\documentclass[acmsmall]{acmart}
\usepackage{array}
\usepackage[utf8]{inputenc}
\usepackage{graphicx}
\usepackage{pdfpages}
\usepackage{hyperref}

\AtBeginDocument{%
  \providecommand\BibTeX{{%
    \normalfont B\kern-0.5em{\scshape i\kern-0.25em b}\kern-0.8em\TeX}}}

\copyrightyear{2024} 
\acmYear{2024} 
\setcopyright{acmlicensed}\acmConference[FAccT '24]{The 2024 ACM Conference on Fairness, Accountability, and Transparency}{June 3--6, 2024}{Rio de Janeiro, Brazil}
\acmBooktitle{The 2024 ACM Conference on Fairness, Accountability, and Transparency (FAccT '24), June 3--6, 2024, Rio de Janeiro, Brazil}
\acmDOI{10.1145/3630106.3658955}
\acmISBN{979-8-4007-0450-5/24/06}

\begin{document}

\title[Machine Learning Data Practices through a Data Curation Lens]{Machine Learning Data Practices through a Data Curation Lens: An Evaluation Framework}

\author{Eshta Bhardwaj}
\email{eshta.bhardwaj@mail.utoronto.ca}
\authornotemark[1]
\orcid{}
\affiliation{%
  \institution{University of Toronto}
  \streetaddress{}
  \city{Toronto}
  \state{Ontario}
  \country{Canada}
  \postcode{}
}
\author{Harshit Gujral}
\email{harshit.gujral@mail.utoronto.ca}
\orcid{}
\affiliation{%
  \institution{University of Toronto}
  \streetaddress{}
  \city{Toronto}
  \state{Ontario}
  \country{Canada}
  \postcode{}
}
\author{Siyi Wu}
\email{reyna.wu@mail.utoronto.ca}
\orcid{}
\affiliation{%
  \institution{University of Toronto}
  \streetaddress{}
  \city{Toronto}
  \state{Ontario}
  \country{Canada}
  \postcode{}
}
\author{Ciara Zogheib}
\email{ciara.zogheib@mail.utoronto.ca}
\orcid{}
\affiliation{%
  \institution{University of Toronto}
  \streetaddress{}
  \city{Toronto}
  \state{Ontario}
  \country{Canada}
  \postcode{}
}
\author{Tegan Maharaj}
\email{tegan.maharaj@utoronto.ca}
\orcid{}
\affiliation{%
  \institution{University of Toronto}
  \streetaddress{}
  \city{Toronto}
  \state{Ontario}
  \country{Canada}
  \postcode{}
}
\author{Christoph Becker}
\email{christoph.becker@utoronto.ca}
\orcid{0000-0002-8364-0593}
\affiliation{%
  \institution{University of Toronto}
  \streetaddress{}
  \city{Toronto}
  \state{Ontario}
  \country{Canada}
  \postcode{}
}

\renewcommand{\shortauthors}{Bhardwaj et al.}

\begin{abstract}
    Studies of dataset development in machine learning call for greater attention to the data practices that make model development possible and shape its outcomes. Many argue that the adoption of theory and practices from archives and data curation fields can support greater fairness, accountability, transparency, and more ethical machine learning. In response, this paper examines data practices in machine learning dataset development through the lens of data curation. We evaluate data practices in machine learning \textit{as} data curation practices. To do so, we develop a framework for evaluating machine learning datasets using data curation concepts and principles through a rubric. Through a mixed-methods analysis of evaluation results for 25 ML datasets, we study the feasibility of data curation principles to be adopted for machine learning data work in practice and explore how data curation is currently performed. We find that researchers in machine learning, which often emphasizes model development, struggle to apply standard data curation principles. Our findings illustrate difficulties at the intersection of these fields, such as evaluating dimensions that have shared terms in both fields but non-shared meanings, a high degree of interpretative flexibility in adapting concepts without prescriptive restrictions, obstacles in limiting the depth of data curation expertise needed to apply the rubric, and challenges in scoping the extent of documentation dataset creators are responsible for. We propose ways to address these challenges and develop an overall framework for evaluation that outlines how data curation concepts and methods can inform machine learning data practices.
\end{abstract}



\maketitle

\section{Introduction} \label{introduction}

The pervasive usage of predictive machine learning (ML) models has not dwindled in the face of ever-growing research discussing cases of biased results \cite{acemoglu_artificial_2022, arora_empirical_2021, bawack_artificial_2022, bi_accurate_2023, davenport_potential_2019, duan_artificial_2019, duchesne_recent_2020, garg_review_2021, henrique_literature_2019, kelly_financial_2023, lam_learning_2023, loureiro_exploring_2018, miotto_deep_2018, peet_machine_2022, kaltenborn_climateset_2023, yao_machine_2023, rolnick_tackling_2022}. Bias in ML models often causes discriminatory, unfair, or unethical judgements towards specific populations. Past research has shown that algorithms can generate gendered biases such as image captioning models that produce gender-specific predictions based on image context \cite{hendricks_women_2018}, analogy generators that associate genders with stereotypical activities \cite{bolukbasi_man_2016}, and neural machine translation systems that generate gendered outputs \cite{tomalin_practical_2021}. Algorithms can also produce racial biases in facial recognition \cite{buolamwini_gender_2018}, inaccurate classifications of racial minorities as ``hateful'' in online hate detection \cite{ahmed_tackling_2022}, and prioritization of referrals for complex medical care for white people over black people on average \cite{ledford_millions_2019}. The biases found in these cases and others are widely attributed to the choices made about datasets used for training ML models \cite{geiger_garbage_2020, holstein_improving_2019, li_data-centric_2022}. 

Reused datasets are not always fit for a new model’s intended purpose. Koch et al. show that benchmark datasets created in one task community are used in other communities, which raises the risk of inappropriate usage \cite{koch_reduced_2021}. Paullada et al. discuss similar concerns on the implications of dataset benchmarks that are reused across tasks and the creation of data derivatives that reuse datasets outside their original context \cite{paullada_data_2021}. Appropriate data use is also hindered by the hidden, tacit, and undervalued nature of the practices underlying data collection, processing, and implementation. As Hutchinson et al. point out, ``How can AI systems be trusted when the processes that generate their development data are so poorly understood?'' \citep[p.~560]{hutchinson_towards_2021}. In addition, knowledge related to using and forming data is often obfuscated because of the tacit skills and expertise involved \cite{muller_how_2019,thomer_craft_2022} but also because data work is undervalued and taken for granted in the face of performance metrics related to models \cite{bender_dangers_2021,borgman_big_2017,hutchinson_towards_2021}. These factors contribute greatly to the lack of transparency and accountability in ML models.

Attempts to address these issues look towards the study of data practices in ML. Data practices in this context are defined as ``…what and how data are collected, managed, used, interpreted, reused, deposited, curated, and so on…'' \citep[p.~55]{borgman_big_2017}, and are also referred to as data work \cite{sambasivan_everyone_2021} and dataset development \cite{khan_subjects_2022,paullada_data_2021,scheuerman_datasets_2021}. Many studies have highlighted that the overall lifecycle for dataset development should get greater recognition for its impact on predictive models and as a result requires a more intentional strategy \cite{bender_dangers_2021,heger_understanding_2022,hutchinson_towards_2021,khan_subjects_2022,paritosh_missing_2018,paullada_data_2021,peng_mitigating_2021,scheuerman_datasets_2021}. This has led to a greater focus on the development of context documents – ``interventions designed to accompany a dataset or ML model, allowing builders to communicate with users'' \citep[p.~2]{boyd_datasheets_2021}. Other research on dataset development has explored the needs of practitioners in performing documentation \cite{heger_understanding_2022,holstein_improving_2019,koesten_collaborative_2019}, the challenges and opportunities in reducing bias and increasing fairness and accountability of data used in ML \cite{akter_algorithmic_2021,liang_advances_2022,mehrabi_survey_2021,miceli_studying_2022,suresh_framework_2021}, the impacts of data preprocessing on ML models \cite{biswas_fair_2021,gonzalez_zelaya_towards_2019,lucchesi_smallset_2022}, aspects of fairness in dataset annotation \cite{koesten_collaborative_2019}, and many more. Particularly, this study adds to emerging research that discusses the adoption of principles from archival studies and digital curation into dataset development processes for machine learning research (MLR) \cite{bender_dangers_2021,colavizza_archives_2022,jo_lessons_2020,leavy_ethical_2021,thylstrup_ethics_2022}.

Digital curation is defined as ``the active involvement of information professionals in the management, including the preservation, of digital data for future use'' \citep[p.~335]{yakel_digital_2007}. The broader domain of digital curation includes all digital objects. Data curation is a subset of this domain that focuses solely on data objects. Data curation involves ``maintaining and adding value to digital research data for current and future use'' \citep[p.~1]{digital_curation_centre_what_nodate}. Studies call for ethical data curation \cite{leavy_ethical_2021} and methods from archival studies as these fields have long dealt with large amounts of data and concerns of representativeness, ethics, and integrity \cite{colavizza_archives_2022, jo_lessons_2020, thylstrup_ethics_2022}. While these studies propose principles and practices that can be adopted from data curation in theory, there is a gap in applying the concepts within ML to demonstrate their feasibility and usefulness in practice. 

In this work, we present an application of a data curation lens within dataset development in ML to obtain a practical understanding of data practices. We review and consolidate the literature on ML data work documentation and data curation frameworks and leverage these theoretical foundations to study whether data curation can feasibly provide frameworks for improved fairness, accountability, and transparency in ML dataset development. Our overall \textbf{research question} is: {\itshape How should data curation concepts and methods inform ML data practices?} Our aim is to explore, at the intersection of these fields, how ML data practices currently perform data curation and how data curation can be enacted more effectively and rigorously. Our \textbf{working hypothesis} is that data curation frameworks can be effectively used to guide and evaluate data practices in ML. We therefore use data curation frameworks to conceptualize and evaluate existing ML practices {\itshape as data curation}. Our goal is that in the near future, data curation is routinely recognized and rigorously performed as a key part of ML research, including its norms and peer review standards. We present a summary of literature from data curation to establish its importance in ML and use it as a lens for ML. By examining data practices in MLR through the lens of data curation, we aim to contribute to effective dataset development in ML that supports transparent, fair, and accountable ML practices and outcomes. 

To connect the two fields, we designed a toolkit to identify gaps and overlaps. It includes a rubric to evaluate the documentation of the contents of datasets as well as the design decisions made in the process of developing datasets based on criteria adapted from the fields of digital and data curation, library, and archival studies. We applied the rubric on sample datasets from NeurIPS, the Conference on Neural Information Processing Systems, a leading global venue for AI/ML research. The design of the framework therefore moves towards the adoption of data curation principles and concepts by influencing evaluation standards. We analyzed the rubric evaluations to understand the entanglement of data practices in the disciplines and determine the feasibility and relevance of assessing ML data work using data curation perspectives. The process of designing and applying this rubric  revealed strengths and weaknesses of current dataset development but also challenges in adapting principles from a data-focussed field like data curation for the model-focussed field of ML. We present our findings in four themes and discuss the limitations of adapting nuanced, practice-based processes from data curation into ML given their differing field epistemologies. We also present pathways to address the four challenges and make recommendations to further progress interdisciplinarity between the fields.

\section{Background} \label{background}

Below, we first review current practices of data work in machine learning research (Section \ref{dataworkinMLR}) and briefly describe foundational data curation concepts (Section \ref{DCtheory}). We then discuss ML studies that start to bridge the fields of ML and data curation and archival studies (Section \ref{DCinMLR}). Finally, we discuss why and how machine learning can adopt data curation to improve current data practices (Section \ref{whyDC}) by extending current studies' use of data curation concepts.

\subsection{Data Work in Machine Learning Research} \label{dataworkinMLR}

In response to the call for accountability and transparency, the development of context documents became the prevalent method of demonstrating the data work involved in ML research. Datasheets, for example, are now a commonly used documentation framework for describing the contents of datasets and select data design decisions made by the dataset creators \cite{gebru_datasheets_2021}. There are also specific structures of context documents for different types of datasets. For example, data statements for natural language processing (NLP) datasets contain specifications on demographic information about the dataset annotator, quality of the dataset, provenance, etc. \cite{bender_data_2018}. Similarly, AI fairness checklists were developed to aid practitioners by providing a structured framework to identify and address issues within their projects \cite{madaio_co-designing_2020}. Model cards aim to ``standardize ethical practice and reporting'' within ML models \citep[p.~221]{mitchell_model_2019}. Model cards include details about the models, their intended use, impacts of the model on the real-world, evaluation data, details on the training data, and ethical considerations \cite{mitchell_model_2019}. Explainability fact sheets are used for similar documentation but are specifically geared towards the method applied in a predictive model. The fact sheet contains an evaluation of the method’s functional and operational requirements, the criteria used for the evaluation, any security, privacy or other vulnerabilities that may be introduced by the method, and the results of this evaluation \cite{sokol_explainability_2020}. 

Simultaneously, dataset development research, sometimes referred to as data science work in ML, became a focal subject of study. Prominently many of these works unearthed how extrinsic and intrinsic biases impact the outcomes of ML models. For example, data cascades - ``compounding events causing negative, downstream effects from data issues, that result in technical debt over time'' \citep[p.~5]{sambasivan_everyone_2021} - result from data practices being undervalued, lack of preparedness in handling data quality in high-stakes domains, data being reused out of context, and data scarcity causing potential downstream risks to groups. Documentation of computer vision datasets have also been analyzed to unearth the values that are prioritized by dataset creators and the field in general \cite{scheuerman_datasets_2021}. ``The kinds of data collected, how it is collected, and how it is analyzed all reflect disciplinary and researcher values'' \citep[p.~4]{scheuerman_datasets_2021}. The results showcase that current practices of dataset development in ML prioritize model development over dataset development, efficiency over reflexive and critical curation, the collection of large, diverse datasets over emphasis on the context and circumstances of the data included in the dataset, and advocate for neutrality and impartiality in their data development process as compared to disclosing their positionality and worldviews \cite{scheuerman_datasets_2021}. Types of intrinsic biases that occur in ML projects have also been organized by building a ``forgettance stack'' with types of forgetting that occur throughout the ML pipeline \cite{muller_forgetting_2022}. ``...forgetting in data science can also be harmful or cause violence, not least because our choice of what we deem unimportant enough to forget to improve our memory, impacts on our understanding of histories, data, exploitation, harm, and so on'' \citep[p.~3]{muller_forgetting_2022}. On the other hand, focussing on intrinsic biases is also seen as failing to acknowledge the power dynamics at play in situations \cite{miceli_studying_2022}. By placing the focus on a bias-oriented framing rather than a power-oriented one, there is a loss of awareness of how labour conditions, social processes, and relationships between dataset creators and consumers impact the data bias present within ML models \cite{miceli_studying_2022}. Instead, it is proposed that research must ``...interrogate the set of power relations that inscribe specific forms of knowledge in machine learning datasets'' \citep[p.~9]{miceli_studying_2022}. 

While many of these studies of dataset development discuss ``data curation'', the term is often used generally to discuss data collection \cite{holstein_improving_2019,li_data-centric_2022,madaio_co-designing_2020}. Contrarily, data curation as a field takes an encompassing lifecycle view and considers many data work processes beyond data collection. The relevance of broader data curation studies to ML is rarely recognized, but several studies identify the opportunities in adopting practices from data curation into MLR.

\subsection{Theoretical Framework of Data Curation} \label{DCtheory}

The information fields of archives, records management, and digital curation share principles, practices, challenges, and knowledge frameworks, but also diverge in areas. Data curation has been defined by institutions in varying ways, on occasion coupled with digital curation \cite{noonan_data_2014}. An important synthesis is made between perspectives that see data curation as digital curation, as value-added infrastructure service, and as an object of archival interest \cite{noonan_data_2014}. Data curation can be defined as, ``...the activity of managing data throughout its life cycle; appropriately maintaining its integrity and authenticity; ensuring that it is properly appraised, selected, securely stored, and made accessible; and supporting its usability in subsequent technology environments.'' \citep[p.~203]{noonan_data_2014}. 

The Digital Curation Center’s lifecycle model consists of stages of curation that projects undergo and helps in identifying roles and responsibilities, processes and best practices, standards and policies, and their documentation \cite{higgins_dcc_2008}. The sequential stages of the DCC curation lifecycle model are `conceptualize', `create or receive', `appraise and select', `ingest', `preservation action', `store', `access, use, and reuse', and `transform' \cite{higgins_dcc_2008}. Data curation emphasizes that each stage of curation must be purposeful and attend to stewardship and future use \cite{palmer_foundations_2013}. The focus lends itself towards ``improvement of data products'' and ensuring data is valuable now and in the future \cite{palmer_foundations_2013}. For each stage of curation, technical, legal, ethical, and operational considerations are made.

\subsection{Data Curation in Machine Learning Research} \label{DCinMLR}

The existing body of knowledge in archival studies, data management, and data curation provide opportunities for adoption within dataset development in ML. Some ML studies have recognized this. For example, Jo and Gebru urge, ``By showing the rigor applied to various aspects of the data collection and annotation process in archives, an industry of its own, we hope to convince the ML community that an interdisciplinary subfield should be formed…''  \citep[p.~307]{jo_lessons_2020}.

Archival science offers sophisticated methods of evaluating, filtering, and curating data that require a high degree of supervision and intervention. While this poses a challenge in some subfields of ML, lessons  can be learned from archives around current key issues in ML including consent, inclusivity, power, transparency, and ethics \cite{jo_lessons_2020}. For example, archives have codes of conduct and ethics to ensure violations do not occur and data curators consider and document whether data should be collected at all based on potential risks and benefits ensuring transparency and supervision of collected data. Leavy et al. further emphasize the importance of ``... [enabling] critical reflection and responsibility for the potential effects of the use of data'' \citep[p.~695]{leavy_ethical_2021}. Their proposed framework for ethical curation consists of 4 principles detailing how to examine the power dynamics of whose voices, labour, and perspectives are included in data curation, how to consider the context and situatedness of data, how to recognize that data curation is a continuous and reflexive process, and how to question the forms of knowledge that are considered legitimate and are included in the data curation process as compared to those that are not. 

Similar to Jo and Gebru, who point out the need for interventions in ML data, Bender et al. describe the risk of documentation debt due to large amounts of uncurated and undocumented data that is used to train large language models \cite{bender_dangers_2021}. The lack of accountability and transparency lead to encoded bias in the datasets used for training. In turn, Bender et al. recommend ``making time … for doing careful data curation and documentation, for engaging with stakeholders early in the design process…'' \citep[p.~619]{bender_dangers_2021}. 

Research at the intersection of archives and ML often focuses on how algorithms can automate archival processes such as extraction, indexing and retrieval, appraisal, and redaction \cite{colavizza_archives_2022}, but some emphasize ``...the opportunity for recordkeeping contributions to the advancement and appropriate use of AI by bringing expertise on provenance, appraisal, contextualisation, transparency, and accountability to the world of data'' \citep[p.~11]{colavizza_archives_2022}. A critical archival approach is required towards datasets in AI to enable reflection on ethical issues such as access, consent, traceability, and accountability \cite{thylstrup_ethics_2022}.

\subsection{How Can Data Curation Benefit ML Data Work?} \label{whyDC}

The ML model development pipeline consists of data collection, data processing, model building, training, model evaluation, and model deployment \cite{hapke_building_2020,paleyes_challenges_2023}. Data curation has similar stages in its lifecycle model. For example, `create or receive', `appraise and select', and `ingest' relate to data collection in ML, while `transform' can involve data cleaning, data augmentation, and data wrangling in ML. However, data curation prioritizes two key aspects within the lifecycle that make it distinct from how dataset development is performed in ML. 

First, data curation has defined inputs, outputs, outcomes, tasks, and reasons for performing each stage in the lifecycle. Importantly, all of these elements are defined and implemented through policies that hold curators and involved stakeholders accountable while also \textbf{enabling transparency}. The `appraise and select' stage evaluates which data should be retained versus discarded for long-term curation. This process is interventionist and requires curators to make judgements on the benefits and risks of storing or discarding the data. In contrast, this is currently missing in ML dataset development where many subfields are driven by collecting the largest amount of data possible. In fact, ML publications introducing new datasets consider the size of the data collected an important contribution when discussing their work. On the other hand, the `appraise and select' stage is performed for 5 reasons: to reduce the amount of data to be curated, to enable efficient retrieval, to enable timely preservation activities, to limit cost of data storage, and to capture legalities of data storage and access \cite{pryor_lifecycle_2012}. The tasks performed in this stage are documented through an appraisal policy which structures the process of making appraisal decisions among other agreed upon requirements for accessibility, retention, etc. The appraisal policy also supports the collection development policy which is an outcome of the prior stage, `receive'. In the next stage, `ingest', in which data is submitted for curation, the appraisal schedule is determined to ensure that there is timely reappraisal of the data being curated to determine needs for further retention and long-term value. These defined guidelines and expectations from each stage of the curation lifecycle enable reuse due to comprehensive documentation, the establishment of clear context and purpose for data curation, and high level of intervention that decreases the risk of introducing intrinsic bias and increases the likelihood of removing or addressing extrinsic bias. Similar standards and processes can be adopted into ML dataset development. 

Secondly, data curation takes a \textbf{lifecycle approach} focusing on adding and maintaining long-term value across each stage, which is reflected in the norms, standards, and practices of data curation communities. The inclusion of stages like `preservation' and `access, use, and reuse' centralizes these reuse-oriented concerns in data curation. These concepts are considered not solely within their specified stages but throughout the dataset lifecycle. For example, considerations around long-term access inform the `conceptualize' stage and data management methods throughout the lifecycle, the `receive' stage identifies access and reuse rights, and the `ingest' stage considers legal ownership issues. The data curation lens therefore not only provides standards and practices but also highlights the value of a cyclical view. 

Pennock outlines the benefits of a lifecycle approach for digital curation, stating that digital materials change throughout their curation process and adopting a lifecycle model facilitates its continuous management \cite{pennock_digital_2007}. This continuity lends itself to the ability to retain authenticity and integrity. A study of data curation at the ICPSR find that data work is often thought of as sequential and is represented through a pipeline but in actuality ``data curation … is a highly collaborative process occurring across a distributed system over time'' \citep[p.~20]{thomer_craft_2022}. 

Data curation supports greater reflexivity on the importance of each stage of data work. It highlights that data reuse now and in the future is dependent on a holistic approach for creating more transparent and accountable datasets which is only possible through meaningful dataset development. In the next section, we discuss the development of a resource that is aimed towards enabling critical dataset development in ML through a data curation lens. 

\section{Methods} \label{methods}

Below, we demonstrate how data curation concepts can be adapted, translated, and operationalized for ML data work.

\begin{figure*}[h]
    \centering
    \includegraphics[width=0.87\linewidth]{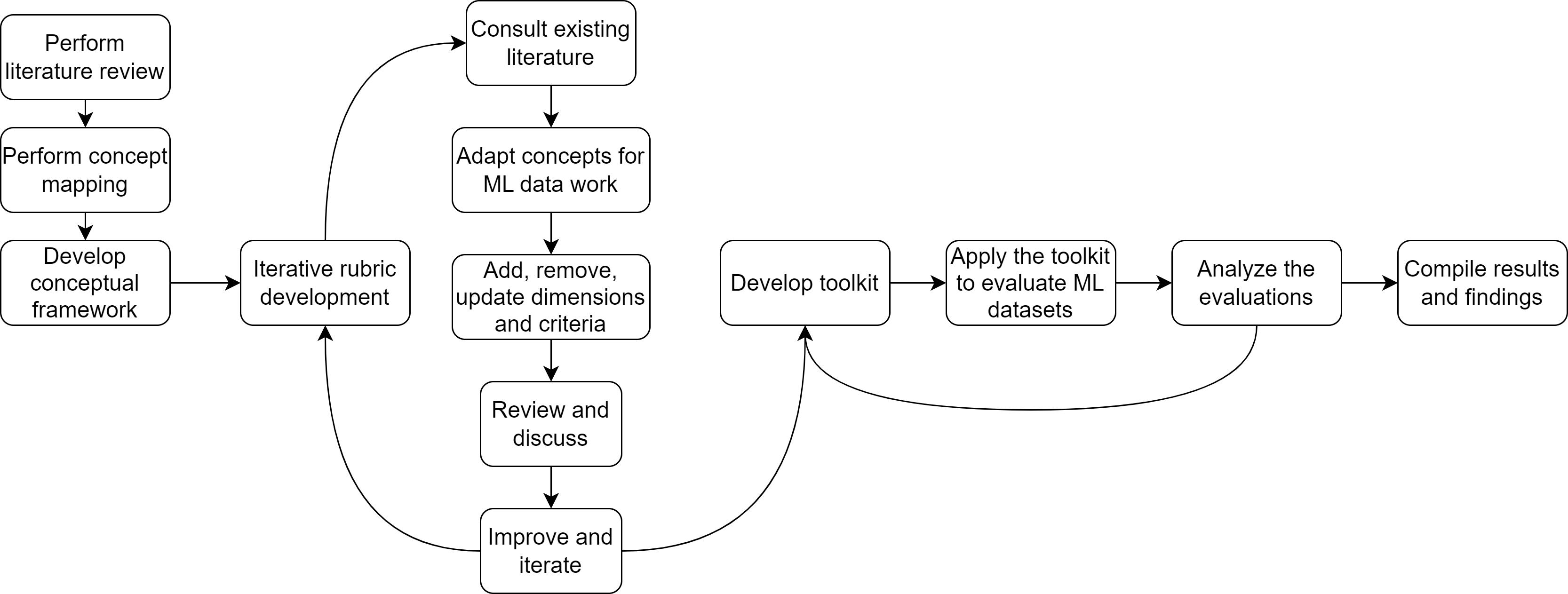}
    \caption{Multi-stage development and evaluation process of the rubric and toolkit}
    \Description{Multi-stage development and evaluation process of the rubric and toolkit}
    \label{fig:fig1}
    \vspace{-0.3cm}
\end{figure*}

\subsection{Development Process} \label{development}

Our framework for evaluating ML datasets centers on a rubric developed in a multi-stage design pictured in Fig. \ref{fig:fig1}. We started by identifying aspects of data curation currently used in ML dataset creation and those that can be further informed by data curation frameworks. Based on concept mapping between the two disciplines supported through literature reviews, we organized dimensions of data curation concepts and principles relevant to ML. We developed the rubric iteratively based on existing literature from digital curation lifecycle models \cite{higgins_dcc_2008}, FAIR data principles \cite{wilkinson_fair_2016}, considerations of environmental sustainability and justice \cite{becker_insolvent_2023,rakova_algorithms_2023}, prior work on digital curation assessment frameworks \cite{becker_design_2020}, and current ML documentation frameworks. The framework builds on the significant impact of datasheets \cite{gebru_datasheets_2021} and takes the logical next step. Datasheets \cite{gebru_datasheets_2021} are focused largely on the content of datasets. Our rubric prompts dataset creators to adopt a reflexive stance about their curation decisions. The earliest drafts of the rubric went through an internal review process in which the authors iteratively discussed and improved the descriptions and evaluation criteria. This included adding, removing, splitting and grouping elements, narrowing down the data quality dimensions most apt for ML datasets, exploring qualitative and quantitative evaluation metrics, and arriving at two levels of evaluation, namely a minimum standard and a standard of excellence. After several iterations of the conceptual framework, we developed additional resources to support the use of the rubric, packaged together as a toolkit.  

We used the toolkit to evaluate select datasets published in the NeurIPS benchmarks and datasets track \cite{yeung_vanschoren_2021}. We collected quantitative and qualitative results on the ratings and comments to understand how data curation is performed in ML, whether data curation principles were effectively adapted for ML datasets to enable feasible evaluations, whether there were elements that emerged as being irrelevant to evaluating ML datasets, and to study the reviewers’ experience, feedback and reflections from applying the rubric. The ratings and results contributed to iterative revisions of the toolkit. In addition, in the final set of evaluations, the reviewers re-examined the evaluations performed for each dataset, and asynchronously resolved disagreements in the ratings by providing comments on whether they agreed or disagreed with another review and accordingly updating their evaluation rating. The reviewers collaboratively discussed the remaining disagreements which helped in further refinement of the rubric and toolkit. In the following sections, we outline the contents of the rubric and toolkit. In Section \ref{findings}, we present our observations and findings from using the rubric to evaluate ML datasets. 

\subsection{Rubric} \label{rubric}

The rubric elements assess the documentation of data composition and data design decisions (i.e., data work) in 19 dimensions across five groups. The full rubric is provided in Appendix A. Below, we briefly discuss a few sample elements within each group.

\textbf{Scope} contains the elements `\textit{context, purpose, motivation}' and `\textit{requirements}'. The latter element’s criteria expect 1) a dataset creation plan and 2) considering how problem formulations can introduce intrinsic biases. This echoes data curation’s emphasis on data management plans that are established at the beginning to guide the entire curation process. The rubric contains these elements because establishing the scope is ``...a translation task from a problem in-the-world, into a problem in-the-business, and then into a data science formulation…Each translation step requires additional interpretation into data sources and data formulations, imposing further decisions upon the humans who carry out the work'' \citep[p.~9]{muller_forgetting_2022}. Capturing these decisions through documentation helps unveil the politics and values involved in setting scope \cite{muller_forgetting_2022,passi_problem_2019,pine_politics_2015}. 

There is an emphasis on reflexivity throughout the rubric, such as being intentional and accountable \emph{while} deciding on the purpose for creating a dataset, but a group of elements are centrally concerned with \textbf{ethicality and reflexivity}: `\textit{ethicality}', `\textit{domain knowledge and data practices}', `\textit{context awareness}', and `\textit{environmental footprint}'. The criteria for evaluating `ethicality' includes a discussion of informed consent and weighing benefits and harms of the dataset. The criteria expects dataset creators to demonstrate `context awareness' by looking inwards and considering how their dataset is a non-neutral representation of the real-world impacted by their perspectives, field epistemologies in which their research is situated, and social, political, and historical context \cite{selbst_fairness_2019}. Dataset creators are also asked to document how their `domain knowledge' expertise and `data practices' shape the dataset. Curatorial work requires craft and unstandardized methods: ``...curators organize their work by first developing a gestalt, abstract mental representation of the data to envision what the final released dataset will entail; they then use their judgement and expertise to interpret standards, [and] creatively come up with solutions…'' \citep[p.~13]{thomer_craft_2022}. Documentation of this tacit knowledge makes it explicit which supports informed choices about reuse. This is supported by Heger et al.’s findings which discuss that ML practitioners ``...noted that information that is implicit or tacit is at risk of being lost if it is not documented'' \citep[p.~13]{heger_understanding_2022}. 

Elements that document \textbf{key stages of the ML pipeline} are included in the rubric because they demonstrate the foundation of how the dataset was developed, namely `\textit{data collection}', `\textit{data processing}', and `\textit{data annotation}'. While these elements are familiar to dataset creators, the rubric offers the opportunity to approach these elements from a different perspective. For example, aside from disclosing the data sources from which data was collected, the rubric urges reflection on how choices in `data collection' have embedded interpretative assumptions because the act of selecting data or ``discovering'' data, especially one source over another, is a human, subjective act that involves interpretation \cite{muller_forgetting_2022}. The rubric also prompts for reflexivity \textit{in the process} of `data collection' rather than at its end. It suggests that criteria for selecting data sources should be discussed and decided prior to its collection in an active process of assessing whether data sources fit the criteria. Ultimately this process must be documented so that data reuse is more transparent, similar to collection development policies in data curation. 

The rubric underscores the application of the data curation lens through the elements about \textbf{data quality dimensions}, including `\textit{suitability}', `\textit{representativeness}', `\textit{authenticity}', `\textit{reliability}' and `\textit{integrity}', along with `\textit{structured documentation}'. `Suitability' prompts dataset creators to reflect on whether their dataset aligns with the purpose they established at the start of the dataset development process and whether the quality of the dataset enables the fulfillment of that purpose. `Representativeness' is included to promote awareness of introducing extrinsic biases through data collection. Dataset creators are asked to define the population represented in their dataset and comment on whether a representative sample is included. `Authenticity', `reliability', and `integrity' are inter-related elements but are analytically separate concepts and specifically defined in archival and digital curation fields. An authentic dataset is one that ``is what it purports to be'' \cite{duranti_reliability_1995,duranti_long-term_2005,duranti_interpares_2007,Higgins_2009,poole_how_2015}. This means that the development of the dataset should include discussion of how `authenticity' was established i.e., how the dataset creators verified the origin of the data they collected. Additionally, it should discuss how authenticity is impacted once the collected data is preprocessed and how the now derived dataset will continue to maintain authenticity. Establishing this chain of authenticity ensures that the dataset that is created is based on verified data and the future reuse of the new dataset can also have a claim of authenticity. A reliable dataset is one that is ``capable of standing for the facts to which it attests'' i.e., that the data points reflect what they represent \cite{duranti_reliability_1995}. The rubric prompts the assessment of the maintenance of `reliability' while creating the dataset and how reliability can be maintained once the dataset is reused. A dataset with `integrity' is one where ``the material is complete and unaltered'' \cite{besek_maintaining_2007,cai_challenges_2015,duranti_protection_1996,Higgins_2009,moore_towards_2008}. The rubric prompts evaluators to check whether dataset creators discuss how integrity has been maintained during dataset creation and future management of integrity. Lastly, we include the `structured documentation' element within this category as the rubric prompts evaluators to assess whether a context document was included to provide documentation about the quality of the dataset’s contents. 

To increase transparency and \emph{appropriate} reuse of datasets in ML, the rubric adopts and adapts the widely used \textbf{FAIR principles for data management} \cite{wilkinson_fair_2016}. The FAIR (\textit{findability, accessibility, interoperability, reusability}) principles were first produced to improve the stewardship and management of research datasets but since then have been adopted into numerous disciplines, including AI/ML \cite{artrith_best_2021,folorunso_fair_2022,jha_implementation_2022,logan_review_2023,noy_are_2023}. In the rubric, documentation for each of `findability', `accessibility', `interoperability', and `reusability' is prompted as individual elements with the principles split into minimum standard and standard of excellence based on their importance and relevance for ML datasets \cite{wilkinson_fair_2016}. Inclusion of the principles in the rubric enables increased transparency and reusability while fostering improved collaboration. 

\vspace{-0.15cm}

\subsection{Toolkit} \label{toolkit}

The conceptual framework of the rubric is complemented with 1) instructions detailing how dataset creators can use the rubric to evaluate their own processes and how dataset re-users (or reviewers) can evaluate existing datasets, 2) guiding principles, recommendations, and FAQ to help in evaluating datasets using the rubric, 3) guidance on interpreting the FAIR principles and authenticity, reliability, integrity, and representativeness, 4) a glossary, 5) and sample evaluations. The toolkit is provided in Appendix B. 

The rubric is used to evaluate a minimum standard and a standard of excellence. The former is evaluated on a pass/fail basis, the latter using none/partial/full. The minimum standard criteria relay the expected level of documentation from all ML datasets while the standard of excellence criteria advocates for a high level of criticality and the documentation only receives ``full'' when all sub-criteria are satisfied. The guiding principles, recommendations, and FAQ sections provide overarching suggestions such as how to approach the evaluation of a dataset that has multiple sources of documentation such as the publication, appendix, website, GitHub page, etc. 

Additional guidance is provided for the data quality dimensions `representativeness', `authenticity', `reliability', and `integrity' as these elements must distinctly be evaluated from an archival and digital curation perspective. For example, `representativeness' is related to `reliability' but more closely focussed on whether the dataset accurately represents the overall set of observations or entities that it claims to be a sample of. Similar guidance is provided around the FAIR principles with simplified explanations and links to self assessment tools and checklists based on the FAIR principles. 

\section{Findings} \label{findings}

\subsection{Application} \label{application}

In order to study whether data curation concepts were feasible for ML dataset development in practice, a set of authors with varied exposure to both ML and digital curation fields conducted a sample set of evaluations using datasets published in NeurIPS. Further information about the authors' expertise is discussed in Appendix C.1. The evaluations were conducted in four rounds (training, round 1, round 2, and round 3). 

\begin{figure*}[h!]
    \centering
    \includegraphics[scale=0.62]{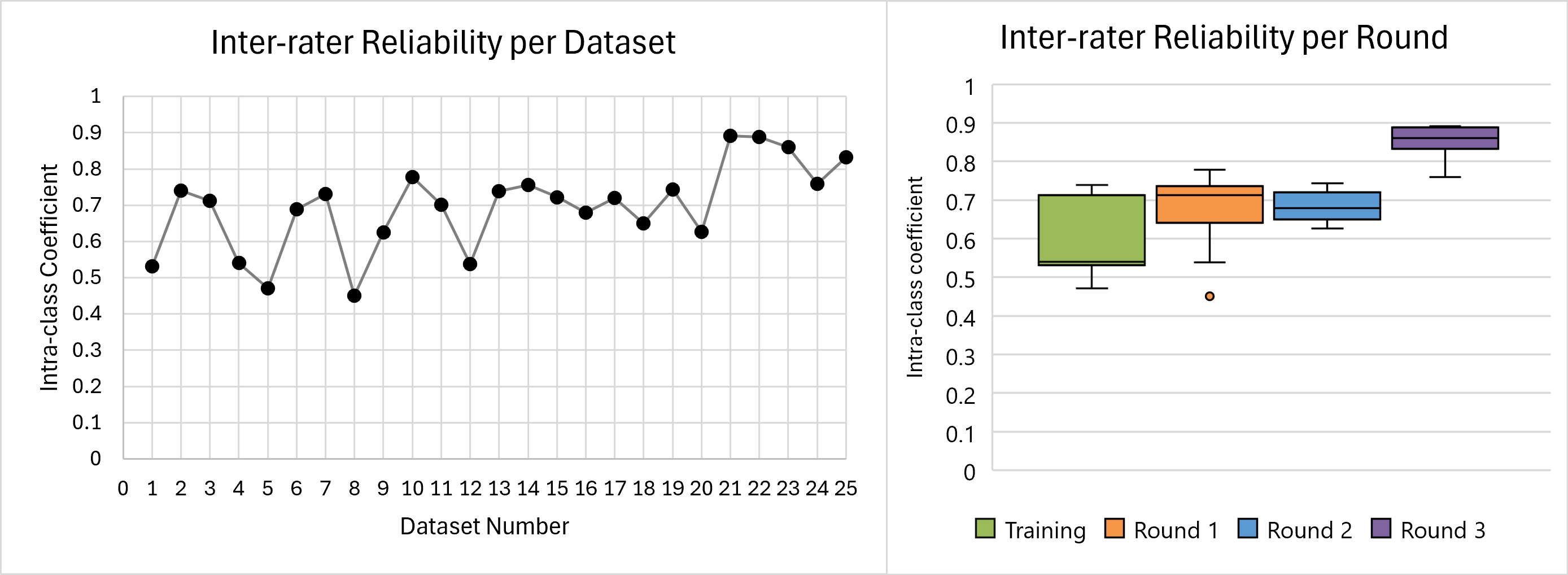}
    \caption{IRR across datasets and rounds}
    \Description{IRR across datasets and rounds}
    \label{fig:fig2}
    \vspace{-0.3cm}
\end{figure*}

We started with a training round so reviewers could become adept with applying the rubric, become familiarized with new concepts and terminology using the toolkit as supplementary material, and ask questions to improve their understanding. The training round consisted of 5 randomly selected datasets published in the NeurIPS benchmarks and datasets track from 2021-2023. Next, three rounds of evaluations were performed on (20) randomly selected datasets; the first round consisted of 10 datasets and the remaining of 5 each. The datasets are listed in Appendix C.2. Following each round, we worked on resolving any disagreements, questions and feedback by improving the rubric and toolkit, and addressing any concerns raised by the reviewers. 

We analyzed the ratings and comments for all the evaluations by measuring inter-rater reliability (IRR). We calculated IRR by using two-way mixed, consistency, average-measures intra-class coefficient (ICC) given our fully crossed design to assess the consistency of the raters’ evaluations of rubric elements across subjects \cite{mcgraw_forming_1996}. Since the ratings for the variables (i.e., rubric elements) were measured on an ordinal scale (i.e., full, partial, none and pass, fail), the ICC was the best suited statistic to assess IRR \cite{hallgren_computing_2012}. ICC values of 1 indicate perfect or complete agreement, 0 indicates random agreement, and negative values indicate systematic disagreement. ICC values of less than 0.40 indicate poor IRR, values between 0.40 and 0.59 indicate fair IRR, values between 0.60 and 0.74 indicate good IRR, and values between 0.75 and 1.0 indicate excellent IRR \cite{cicchetti_guidelines_1994}. 

Fig. \ref{fig:fig2} shows the progression of IRR from training (datasets 1-5), round 1 (datasets 6-15), round 2 (datasets 16-20), and round 3 (datasets 21-25). The lowest ICC value is 0.45 for dataset 8 which indicates fair agreement while the highest is 0.89 for dataset 21 which indicates excellent agreement. Across datasets, the IRR values span from fair to excellent which indicates the difficulty in obtaining truly consistent evaluations. Nonetheless, the distribution of IRR per round shows lower variability in the consistency as the rounds progress indicating gradual improvement through iterations. Furthermore, round 3 has all 5 ICC values indicating excellent agreement (ICCs between 0.76 and 0.89). We also compared reliability across elements, which had mixed results, as discussed further in Appendix C.3. 

To assess the extent to which the rubric and toolkit improvements between the evaluation rounds were impacting the consistency of the evaluations, we analyzed the disagreements by datasets and elements. A summary of the average number of inconsistencies across datasets can be found in Appendix C.4. Most importantly, the metric demonstrates that the overall percentage of all disagreements decreased from training (32\%),round 1 (25\%), round 2 (23\%), to round 3 (7\%) indicating that the iterative development of the rubric was improving the consistency of the evaluations.

We reviewed the inconsistencies across each element of the rubric to determine whether any specific elements were standing out as infeasible to adapt from data curation to ML or required further improvements to adapt. To measure this, we calculated the percentage of datasets with inconsistencies for each rubric element as shown in Appendix C.5. 

We analyzed responses after each round and consequently introduced changes to the toolkit that would reduce inconsistencies iteratively. As a result, we were able to identify improvements in some rubric elements and recurrent challenges in others. Difficulties with curation-specific terminology such as the difference between `findability' and `accessibility' was addressed by clearer definitions and examples. Other difficulties, such as the applicability of data quality evaluation for datasets that were synthetic (not collected) were addressed with better guidance. Lastly, `reliability', `authenticity', and `integrity' were difficult to evaluate because while the documentation provided by dataset creators acknowledged the limitations of the dataset, it was not in terms that addressed these elements specifically. 

To get a better understanding of the reason for the inconsistencies after round 2, we analyzed the evaluations by comparing the evaluation comments against a ``reference'' comment. Based on this analysis, specific patterns emerged on the reasons for disparate evaluations between raters. These reasons, in turn, revealed four types of challenges in applying data curation concepts to ML contexts. A sample set of analyzed evaluations is presented in Table \ref{tab:tab1}, and discussed in further detail in the following section. 

In response to the range of evaluation outcomes for the `structured documentation' dimension, we reviewed the current data practices reported by the dataset creators in more detail by analyzing the context documents provided with publications. Our review, detailed in Appendix C.6, indicates that out of 25 datasets assessed, 6 lacked an accompanying context document. Of the 19 datasets with context documentation, we identified limitations that undermine their completeness and utility. This review highlights instances where modifications to standard datasheets or checklists by dataset creators lead to the omission of essential curation details. We document cases where the provided information was ambiguous or could not be independently verified, emphasizing the need for improved documentation standards to uphold the integrity of data curation processes.

\subsection{Challenges} \label{challenges}

Table \ref{tab:tab1} introduces four challenges we identified through the evaluation results. They illustrate the difficulties of designing an evaluation framework that assesses ML concepts using data curation principles. These challenges are not comprehensive but serve as a demonstration of salient issues in this interdisciplinary space. 

\subsubsection{False Friends}
Some elements refer to terms shared between data curation and ML (or computing broadly) that have non-shared meanings {\itshape (false friends)}. For example, `reliability' in engineering and computing disciplines refers to expected consistency in performance (i.e., that a system will perform as expected in a given time period and environment). For datasets, this is often interpreted as the trustworthiness of data in terms of accuracy and consistency \cite{wang_beyond_1996}. However in data curation, reliability is defined as whether data is ``capable of standing for the facts to which it attests'' \cite{duranti_reliability_1995}. For the example provided in Table \ref{tab:tab1}, raters evaluated the standard of excellence for `reliability' for dataset 19, which has criteria stating that the documentation discusses the management of reliability for appropriate reuse in the future, i.e., how the dataset structure and documentation enable reliable re-purposing and reuse. Interpreting this criteria as ``dataset reliability'' leads to consideration towards whether the dataset would be accurate over time for reuse and consistently available. Accordingly, Rater 3’s evaluation points to a discussion on maintenance and findability of the dataset rather than an evaluation of processes in place to ensure that the dataset will continue to represent the information it is about even when it is reused and repurposed. 

\subsubsection{Interpretative Flexibility}
The rubric's more open ended criteria lead to {\itshape interpretative flexibility}, which can result in divergent ratings. For example, the evaluation of ethicality in dataset 20 surfaced how different standards and expectations can collide, resulting in a full range of evaluations. While one rater was fully satisfied by a discussion of potential negative impacts (full), another recognized these statements as typical but expected more (partial), and the third considered them insufficient (none). 

\begin{table*}
    \centering
    \caption{Sample set of round 2 evaluations and challenges}
    \small 
    \resizebox{\textwidth}{!}{
    \begin{tabular}{|p{0.04\textwidth}|p{0.1\textwidth}|p{0.25\textwidth}|p{0.25\textwidth}|p{0.25\textwidth}|p{0.15\textwidth}|}
        \hline
         Data-set & Element & Evaluation Comments (paraphrased) & Reference Comment & Reason for Inconsistency & Challenge\\
         \hline \hline
         19 &  Reliability, standard of excellence & Rater 2 (none) mentioned that there was no specific discussion of reliability as it pertains to reuse. Rater 1 (partial) pointed to the maintenance section of the datasheet. Rater 3 (full) pointed to a DOI and maintenance plan as assurance for long-term reliability. & I would rate this as none. Despite the maintenance section in the datasheet, the response does not discuss maintenance as it pertains to maintaining reliability when the dataset is repurposed and reused. & Reliability is interpreted from a software or computing perspective which considers consistent performance rather than a data curation perspective which considers how data will remain true to the facts it represents through reuse. & False friends\\
         \hline
         
  20 & Ethicality, standard of excellence & Rater 1 (none) stated that the documentation doesn’t go beyond standard ethics statements. Rater 2 (partial) stated that documentation on potential negative impacts is identified. Rater 3 (none) states that there is no identifiable risk in this dataset. & I would rate this as a none because there is no further discussion of ethics beyond typical negative impacts statements. & Rater 3 interprets this dataset as being as low-risk for ethicality and doesn’t believe there is a need to ``go beyond requirements listed in ethics framings''.  & Interpretative flexibility\\
         \hline
              
         17 & Inter-operability, both levels & Rater 1 gave a fail (minimum) and none (excellent) and mentioned there was no explicit documentation of how the dataset integrates with other workflows. Rater 2 (also fail/none) mentioned that machine and human readability is discussed implicitly because data is in a CSV format but fails for lack of discussion on integration. Rater 3 (pass/full) mentioned all relevant info was given on GitHub. & I would rate this as a pass for minimum standard because data is in a popular, standard format. I would rate this as none for excellence because controlled vocabularies and qualified references for linking were not used/discussed. & It is difficult to decide to which extent human and machine readability should be evaluated. The reference comment indicates that a popular, standard format is sufficient. However, while CSV is a popular format, it can only be processed if all columns are fully defined. The reviewers would need expertise about multiple data formats and their structures to fully assess this. & Depth of analysis\\
         \hline
         19 & Domain knowledge and data practices, minimum standard & Raters 1 and 2 gave a fail because there was no explicit documentation about this element. Rater 3 gave a pass, and mentioned that expertise is required in curation, web crawling, and natural language processing. & I would rate this as a fail, because there is no explicit discussion on how the process of developing this dataset required special skills/expertise. & The documentation describes the curation of LLMs as intensive and specialized (presented as a description of the problem domain). This is however not a description of the knowledges required to develop this dataset. The challenge for the raters is to interpret the extent of documentation the dataset creators are responsible for. & Scoping\\
         \hline
    \end{tabular}}
    \label{tab:tab1}
    \vspace{-0.5cm}
\end{table*}

\subsubsection{Depth of Analysis}
The third challenge arose as a result of reviewers bringing differing expertise and different technical know-how to evaluating an element, but the important question is how deep an evaluation can and should go beyond surface documentation. 
Table \ref{tab:tab1} points to an example of this for evaluating `interoperability' for dataset 17. The criteria direct reviewers to assess whether the metadata and data are readable by humans and machines. This can be interpreted by evaluating whether the dataset is made available in a standardized and documented format. Data format standardization however has multiple levels. For example, even a structurally simple standardized `container' format such as CSV must be complemented with clear definitions of each column. For more complex data, the recursive analysis and exhaustive models of dependency networks can become effort intensive \cite{giaretta_advanced_2011}.

\subsubsection{Scoping}
The last challenge in designing the rubric was {\itshape scoping} the expected standard of documentation from dataset creators. For example, in evaluating the maintenance of integrity while developing the dataset (minimum standard) and management of integrity for appropriate reuse in the future (standard of excellence), it is challenging to scope which points in the data pipeline the dataset creators are responsible for documenting processes around integrity. In the example of dataset 16, raters reported confusion around whether the integrity should be evaluated based on the integrity of the collected data, or the maintenance of integrity in the data pipeline, or the integrity of the final produced dataset. Similarly, in the example shown in Table \ref{tab:tab1} of dataset 19, `domain knowledge and data practices' was challenging to evaluate because it was unclear whether expertise in collecting the data, the problem domain overall, or developing the dataset needed to be documented. 

\section{Discussion} \label{discussion}

\subsection{Limitations} \label{limitations}
We identify two key limitations in the application of the rubric. First, using the rubric to evaluate ML datasets requires training, practice, and familiarity with data curation concepts. Performing evaluations iteratively and taking part in workshops and discussions help improve the required data curation knowledge. This also creates a potential scenario in which ML experts may be expected to acquire an unreasonable amount of expertise in data curation prior to applying the rubric. Uptake of such a rubric requiring specialized knowledge and skills that are improved over time is presently a limitation on the immediate resolution of using data curation to improve fairness, accountability, and transparency in ML dataset development. 

Second, our current evaluation framework is used to explore the connections between data curation and ML dataset development through its application on a select set of datasets where evaluations are performed by a select set of reviewers. In addition, the reviewers were trained in using the rubric. Furthermore, the results from the application of the rubric are on the basis of randomly selected datasets that {\itshape aim} to represent ML datasets at large. This means that our findings are predicated on these factors. This further implies that the improvements made to the toolkit are on the basis of difficulties faced in evaluating a sample set of datasets. We report IRR metrics that showcased improved consistency in responses between each round. However, we cannot distinguish to what degree the ICC values improve because the toolkit was updated and improved after each round, or because the reviewers became more consistent at interpreting and evaluating the rubric criteria. 

\subsection{Pathways Forward} \label{pathways}
We outline some recommendations to address the challenges based on the lessons learned from applying a data curation lens to examine ML data practices. These challenges combine problems that can be fixed with tensions that will remain present and need navigation, thus they present opportunities for growth between the disciplines through continued exploration of the intersections in data practices, including further toolkit development. 

The presence of {\itshape false friends} across fields suggests that evolving documentation can aid with defining, understanding, and navigating the differences in shared terms. Toolkit components like the glossary and FAQ can provide evolving required context to ensure that shared terms between ML and data curation are evaluated as intended. 

The challenge of {\itshape interpretative flexibility} presents an opportunity to engage in generative discussion and collaboration that broadens the association between data curation and ML dataset development. As with any form of descriptive evaluation, the rubric necessitates interpretation. The recent emergence of research intersecting these fields means that evaluation across the disciplines is complex. One consideration for generative discussion is to what extent (and if it all) the evaluators need to agree in their assessments. It can be argued that a better approach would be to embrace the flexibility of the evaluations within the format of the rubric and create an evaluation framework that doesn’t result in ratings and comments but questions and recommendations to foster collaboration instead of dissonance. In fact, one of the identified challenges of enforcing triangulation is that it acts as a barrier to collaboration \cite{archibald_investigator_2016}. Instead, approaching interpretative disagreements as a way to understand evaluators’ perspectives can prompt deeper reflexivity \cite{archibald_investigator_2016,johnstone_weighing_2007}. This can be especially helpful in progressing the interdisciplinarity between the fields at this early stage of intersection. 

The challenge of {\itshape depth of analysis} is linked to {\itshape interpretative flexibility} because requiring agreement in evaluations means requiring identical levels and types of expertise from the reviewers. In other words, the optimal depth of analysis will vary because depth of expertise varies and disagreements happen on different levels. However, as we discussed above, if the evaluation framework does not require agreement among reviewers, the disagreements arising due to {\itshape depth of analysis} can become prompts for deeper levels of assessment. Disagreements would then become generative and would be used as a starting point for discussion. 

The related challenge of {\itshape scoping} occurs due to inevitable entanglement of data curation and ML. The processes of curation and dataset development are inseparable in practice, yet conceptually separable even when occurring contemporaneously. Setting clear boundaries on the expectations from data creators can aid in scoping the documentation they are responsible for. But as datasets regularly reuse prior datasets, it is not easy to determine the appropriate boundary of responsibility for the quality of data curation. A guiding principle for this boundary, adopted from data curation, can be to maintain the chain of custody, i.e., dataset creators should be expected to provide all possible documentation relating to their processes and pointing to others’ documentation for processes outside their control. 

\section{Conclusion} \label{conclusion}

Jo and Gebru ``hope[d] to convince the ML community that an interdisciplinary subfield should be formed…''  \citep[p.~307]{jo_lessons_2020}. In order to make sense of the intersecting terminologies and concepts in this interdisciplinary space, we must develop the right tools. Here, we explore what form and content these tools might take. The paper explored the intertwined relationship of data practices in data curation and ML and presented a method for how data curation concepts can be adapted for ML dataset development. The process of exploring this intersection of fields yielded a high-level framework of dimensions and criteria as well as insights into the challenges of merging these fields’ perspectives. We adopted standards for transparency and accountability built into data curation processes to evaluate the documentation of dataset development in ML. 

Based on our data, we claim that the evaluation enabled by the framework identifies strengths and weaknesses in order to prioritize targeted improvements by incorporating data curation methods where they are most needed. As a diagnostic aid, the formative evaluation helps ML practitioners decide how to improve their dataset’s documentation and develop staged objectives to improve their practices. Aggregate evaluation results highlight priorities, such as environmental footprint disclosures. By incorporating data curation norms, evaluation criteria, and terminology into evaluation guidelines for ML, the framework contributes to normalizing the idea that data curation is part of ML and guides the community in systematically addressing and evaluating it.

This work answers calls for data curation in AI/ML \cite{colavizza_archives_2022,jo_lessons_2020}, supports the examination of intrinsic and extrinsic biases in the dataset development process, and facilitates greater reflexivity \cite{leavy_ethical_2021}. Our results demonstrate the potential of collaboration between data curation and ML data work, with the toolkit as a resource for bridging the gap in practice.

\begin{acks}
This research was partially supported by NSERC through RGPIN-2016-06640 and the Canada Foundation for Innovation.
\end{acks}

\nocite{dignazio_data_2023,muller_designing_2021,julian_posada_platform_2023,lin_trust_2020,klein_scholarly_2014,zenodo_-_research_shared_fair_nodate,go_fair_i3_2017, shen_multi-lexsum_2022,chen_dataset_2021,fiesler_participant_2018, information_and_privacy_commissioner_of_ontario_consent_nodate, go_fair_f1_2017,liang_reflexivity_2021,bardzell_towards_2011,mcintyre_doctrine_2023,fair_principles_r12_nodate,sweet_who_2020,digital_curation_centre_glossary_nodate,kim_martineau_what_2021,leenings_recommendations_2022,matthew_stewart_olympics_2023,ruiz_learning_2019,schuster_programming_2021,saito_open_2021,moosavi_scigen_2021,luo_moma-lrg_2022,hormazabal_cede_2022,wang_loveda_2021,aakerberg_rellisur_2021,hendrycks_measuring_2021,huang_dgraph_2022,mall_change_2022,islam_caesar_2022,xu_globem_2022, kaltenborn_climateset_2023, hassan_bubbleml_2023,gadre_datacomp_2023,horel_cpd_2021,dell_amerstories_2023,lee_visalign_2023,hambro_dungeons_2022,larson_evaluating_2022,stanley_fsmol_2021,huang_tufts_2021,mazeika_how_2022,penedo_refinedweb_2023,kuang_stanford-orb_2023}

\bibliographystyle{ACM-Reference-Format}
\bibliography{firsthalf_DCML_FAccT24,secondhalf_DCML_FAccT24}



\section*{Supplementary material}

**A    RUBRIC DOCUMENTS**

**A.1    Rubric**

| | CURATORIAL ELEMENT | DESCRIPTION | DOCUMENTATION LEVEL | |
|---|---|---|---|---|
| | | | Criteria to meet minimum standard | Criteria to meet standard of excellence |
| | | | SCOPE | |
| 1 | Context, purpose, motivation | This information explains the purpose of dataset creation for the specified domain. | Documentation discusses the problem domain, what problems the new dataset addresses, the relevance of those problems, and the need for a new dataset in comparison to existing datasets. | Documentation states how the dataset can be reused beyond its original context. |
| 2 | Requirements | The translation process from a "real-world" problem to a "ML problem" for which the dataset is created [105,112] consists of numerous decisions, expertise, and worldviews that should be documented in order to understand the context in which the problem situation was framed. | Documentation states how the problem was formulated and how the dataset creation plan was generated. | Documentation includes reflection on how the problem formulation introduces intrinsic biases and states different approaches in formulating the problem apart from the final presented plan. |
| | | | ETHICALITY AND REFLEXIVITY | |
| 3 | Ethicality | Ethical considerations are critical to the fair and accountable creation and (re)use of datasets. | Documentation discusses how the benefits of creating the dataset outweigh any harms of creating it (see proportionality principle), and it discusses informed consent if the dataset is about humans. | Documentation goes beyond requirements listed in ethics framings like guidelines/policies/checklists. For example, documentation discusses alternate methods of dataset creation that were not used because of potential ethical harm. |
| 4 | Domain knowledge & data practices | Creating a dataset involves, often tacit, expertise about one or more domains as well as data practices. Articulating both types of nuance required in dataset development makes data work more transparent [51,67,105,116,133]. | Documentation discusses if developing the dataset required expertise in a given domain and specific data skills. | Documentation discusses whether specific skill-set/expertise is required for reusing the dataset. |
| 5 | Context awareness | Context awareness demonstrates an understanding of the subjective, non-neutral nature, and situatedness of data. | Documentation includes a positionality statement or similar reflection on the dataset creators' awareness of social, political, and historical context. | Documentation adopts a reflexive approach to dataset development. For example, documentation discusses how field epistemologies impact assumptions, methods, or framings. |
| 6 | Environmental footprint | This element is for dataset creators to reflect and quantify the footprint of their dataset creation process [13]. | Documentation contains a quantitative assessment of environmental footprint and clearly defined scope of what was measured. | Documentation includes a lifecycle assessment and the corresponding environmental footprint, and an assessment of design choices and rationale for the choices. |
| | | | DATA PIPELINE | |
| 7 | Data collection | Disclosing data sources is essential in the data collection process. Further reflection on the process of selecting those sources can reveal important interpretive assumptions [105] and historical and representational biases [67]. | Documentation states how and why data was gathered from the data source(s). | Documentation discusses the process of defining criteria for selecting data sources, specifies the criteria, explains why those criteria were chosen, and how the selected data sources are validated against these criteria. |
| 8 | Data processing | Data processing involves cleaning, transforming, and wrangling data. Data processing decisions have impacts on the ultimate "cleaned" data that is used [90,105]. Detailed documentation of this process enables outcomes of the model to be traced back to processing decisions. | Documentation discusses the process of cleaning, transforming, or wrangling data. | Documentation goes beyond what is done to discuss how the decisions about data processing were made and why, and potential impacts of the processing decisions. |

| 9 | Data annotation | Data annotation or labelling, regardless of the guidelines provided to reduce worker bias, can lead to disagreements on how data should be annotated (either between annotators or between dataset creators and annotators).The inclusion of this documentation highlights what is considered the "ground truth" [29,105,106] by the dataset creators which impacts how annotation is performed [69]. | Documentation discusses the process of annotation, and if any labels are used, the documentation includes the following:<br><br>If data is synthetic: documentation states whether data is generated to match labels and how the relationship of the data to the labels is verified.<br><br>If data is collected, documentation states how data was interpreted to generate labels. | If data is synthetic: documentation discusses how annotations were generated to be robust, i.e., not sensitive to variability.<br><br>If data is collected: documentation discusses how disagreements on annotation are reconciled and if labels are used, documentation includes reflection on how labels represent differing worldviews and social backgrounds. |
|---|---|---|---|---|
| | | | **DATA QUALITY** | |
| 10 | Suitability | Suitability is a measure of a dataset's quality with regards to the purpose defined. | Documentation discusses how the dataset is appropriate for the defined purpose. | Documentation discusses how dimensions such as accuracy, completeness, timeliness, and consistency contribute to the quality of the dataset in being used for the defined purpose. For example, timeliness (i.e., age) of data should be appropriate for the defined purpose. |
| 11 | Representativeness | Representativeness is a measure of how well a sample set of data represents the entire population. Sampling procedures and decisions about data sources can introduce extrinsic bias [105]. For example, choosing Reddit or Twitter as a data source can perpetuate dominant social biases rather than being a representative sample of the target population [13]. | Documentation defines the population and discusses the extent to which the sampling procedure is representative of the population. | Documentation includes reflection on how the dataset creation process overall, and the sampling procedures specifically, affect extrinsic bias. |
| 12 | Authenticity | Authenticity of a dataset is about whether the dataset "is what it purports to be" [32-34,56,119], which is a responsibility of dataset creators [87]. | Documentation discusses how authenticity has been established and maintained including how the dataset creators verified the origin of all data they use and how data processing impacted authenticity. | Documentation states how the dataset will maintain authenticity in the future, i.e., preservation processes in place to ensure that future reuse of the dataset can also have a claim of authenticity. |
| 13 | Reliability | Reliability is about how the dataset is "capable of standing for the facts to which it attests" [32], i.e., how its data points reliably reflect what they represent. | Documentation discusses how the reliability of the dataset has been maintained, including the verification and validation steps taken to ensure reliability, where necessary. | Documentation discusses the management of reliability for appropriate reuse in the future, i.e., how the dataset structure and documentation enable reliable re-purposing and reuse. |
| 14 | Integrity | Integrity of a dataset is about whether "the material is complete and unaltered" [14,21,35,56,102]. | Documentation discusses how the integrity of the dataset has been maintained. | Documentation discusses the management of integrity for appropriate reuse in the future, i.e., preservation processes in place to ensure accuracy and consistency over time. |
| 15 | Structured documentation | Context documents in standardized structures provide information on the content of the dataset which is critical in establishing its usage in a well defined format. | Documentation includes a standardized context document. Acceptable formats include but are not limited to datasheets, data statements, and nutrition labels. | The context document addresses all mandatory items. |
| | | | **DATA MANAGEMENT** | |
| 16 | Findability | Ensuring findability is about enabling the dataset to be discovered for reuse after its development [138]. | Documentation discusses how the dataset is findable by providing a | Documentation includes metadata and the metadata and data are stored in a searchable repository. |

| | | | globally unique and [persistent identifier](URLs are not persistent). | |
|---|---|---|---|---|
| 17 | Accessibility | Accessibility is about enabling the dataset to be obtained after its development [138]. | Documentation states all information and tools required to access the content of the data, and the identifier navigates to the metadata and data. | Documentation includes a communications protocol, an authentication and authorization procedure, and provides metadata that will be available even if data access is removed. |
| 18 | Interoperability | Interoperability ensures that the dataset can be integrated with other applications and workflows [138]. | Documentation discusses how the dataset integrates with other data, workflows, applications, etc. (i.e., that the metadata and data are readable by humans and machines). | Documentation has metadata and data that use controlled vocabularies and link to other resources using qualified references. |
| 19 | Reusability | Ensuring reusability requires providing information such as relevant [provenance](#) and usage [138]. | The metadata and data include provenance information including where the data came from, who collected it, and when. | Documentation has metadata and data that are described using domain-relevant standards, states license and usage information, and provides additional provenance documentation as described by FAIR best practices. |

**A.2      Rubric Worksheet**

| | CURATORIAL ELEMENT | DOCUMENTATION LEVEL | | | |
|---|---|---|---|---|---|
| | | Criteria to meet minimum standard | | Criteria to meet standard of excellence | |
| | | Pass/Fail | Comments | Full/Partial/None | Comments |
| | SCOPE | | | | |
| 1 | Context, purpose, motivation | | | | |
| 2 | Requirements | | | | |
| | ETHICALITY AND REFLEXIVITY | | | | |
| 3 | Ethicality | | | | |
| 4 | Domain knowledge & data practices | | | | |
| 5 | Context awareness | | | | |
| 6 | Environmental footprint | | | | |
| | DATA PIPELINE | | | | |
| 7 | Data collection | | | | |
| 8 | Data processing | | | | |
| 9 | Data annotation | | | | |
| | DATA QUALITY | | | | |
| 10 | Suitability | | | | |
| 11 | Representativeness | | | | |
| 12 | Authenticity | | | | |
| 13 | Reliability | | | | |
| 14 | Integrity | | | | |
| 15 | Structured documentation | | | | |
| | DATA MANAGEMENT | | | | |
| 16 | Findability | | | | |
| 17 | Accessibility | | | | |
| 18 | Interoperability | | | | |
| 19 | Reusability | | | | |

**B        TOOLKIT**

## Overview of Research

**Background and Motivation**: The usage of artificial intelligence has increased exponentially with applications in predicting outcomes related to education, employment, housing, and many more social, economic, and financial aspects of our lives. Archival studies have long dealt with large amounts of data and concerns of representativeness, ethics, integrity, and more with the use of data curation methods, theories, and frameworks. Machine learning research (MLR) has pinpointed the data underlying predictive models to be the largest contributor in introducing bias [113,124,125]. Emerging studies have advocated for the prioritization of rigorous data curation practices often referred to as "data work" or "dataset development" in MLR [13,51,72]. Introducing data curation concepts and principles can therefore improve the transparency and accountability of the dataset creation process within MLR.

**Objectives:** We assess ML dataset development processes using principles and methods from archival studies and digital curation. We perform a synthesis and organization of existing work to enable the coherent usage of data curation frameworks, a taxonomy of data curation terms used within machine learning research, and a review of gaps and opportunities for data curation in machine learning.

**Method**: Our research design for this study consists of the following:

1. Synthesizing literature on data curation concepts and principles central to ML data work.
2. Exploring the relevance of data curation concepts and principles through an illustration of how they can be adapted, translated, and operationalized for ML data work.
3. Demonstrating the gaps and overlaps in how ML data practices already perform data curation, how data curation is discussed in MLR, and how data curation can be further adopted.

**Goals and contributions**: This project deepens the scholarly and practical connections between the data curation and machine learning research communities and initiate directions for improvement within MLR's data practices. The outcomes present a novel perspective on improving documentation practices in machine learning through data curation. Through this project, we aim to further establish the connections between the data curation and machine learning research communities.

## Application Guidance

**Scope of application:** The rubric is intended for two types of users.
1. Firstly, dataset creators can use the rubric as a resource to prompt and facilitate critical engagement and reflection throughout their dataset creation process.
2. Secondly, existing datasets can be evaluated prior to publishing or reuse by applying the rubric to determine gaps that require further documentation and areas where bias can be introduced. In both cases, we aim for the rubric to be a practical and useful resource for researchers to engage with the dataset creation process using a data curation lens. The rubric was developed for the evaluation of ML datasets and has elements specific to the domain, including: requirements, data annotation, environmental footprint, and structured documentation.

**Applying the rubric to your own dataset**

The overall process for using the rubric is as follows:

1. Read the rubric to get familiarized with the elements and details that will be needed.
2. Review each element in the rubric individually.
   a. For each element, first assess whether the minimum standard of documentation has been fulfilled. To do this, provide a pass/fail evaluation, where a pass is granted for *any* amount or type of discussion around the element and a fail is granted *only* if there is no discussion around the element at all.
   b. Next, assess whether the documentation meets a standard of excellence, only if the minimum criteria received a pass. The standard of excellence is a full/partial/none evaluation. A full is granted if all aspects specified in the standard of excellence column were discussed, a partial is granted if one or more (but not all) were discussed, and a fail if none were discussed.
   c. It is important to note both for points 2a and 2b that the quality of the responses/documentation is not being assessed but rather if the element was considered and reflected on in any capacity. The purpose of the rubric is to demonstrate the dataset creators' thought process and provide transparency so that its reuse is based on a complete understanding of the dataset.
3. For each element, along with the grade, a comment on what specific information was used to determine that grade must be provided. Other comments and questions can also be included.

The evaluation of each dataset can take 30-60 minutes.

**Applying the rubric to existing datasets through publications**

The overall process for using the rubric is as follows:

1. Read the rubric to get familiarized with the elements and details that will be needed.
2. Gather and review all pertinent information that can be found about the dataset. This will include the research paper, appendices, the linked dataset, and any documentation associated with the externally linked dataset (e.g., README on GitHub).
3. Review each element in the rubric individually by looking for it across all the information gathered in step 1. Some of the elements will be easier to locate than others because they will be titled specifically, whereas others may be discussed at any point.
    a. For each element, first assess whether the minimum standard of documentation has been fulfilled. To do this, provide a pass/fail evaluation, grant a pass for *any amount* or type of discussion around the element and fail *only* if there is no discussion around the element at all.
    b. Next, assess whether the documentation meets a standard of excellence, only if the minimum criteria received a pass. The standard of excellence is a full/partial/none evaluation. A full is granted if all aspects specified in the standard of excellence column were discussed, a partial is granted if one or more (but not all) were discussed, and a fail if none were discussed.
    c. It is important to note both for points 2a and 2b that the quality of the responses/documentation is not being assessed nor the correctness of the technicalities but rather if the element was considered and reflected on in any capacity. The purpose of the rubric is to demonstrate the dataset creators' thought process and provide transparency so that its reuse is based on a complete understanding of the dataset and how it was developed.
4. For each element, along with the grade, a comment on what specific information was used to determine that grade must be provided. Other comments and questions can also be included.
5. For each dataset, evaluators must provide a reflection on their overall assessment of the documentation and rigour demonstrated in the dataset creation process.
6. For each dataset, evaluators must provide a confidence rating for their evaluation.

We estimate the evaluation of each dataset will take about 30-60 minutes once you are familiar with the framework.

**How to interpret authenticity, reliability, integrity, and representativeness**

It may be worth noting that the archival and digital curation perspectives that inform the evaluation framework are particularly important to interpreting the meaning of certain dimensions. Above all, the cluster of authenticity, integrity and reliability needs to be understood from this angle. They are closely related aspects, often treated or addressed by similar mechanisms, but they can be seen as analytically separate concepts. Here is an example.

When you download a data set of weather observations from a platform, you may want to verify if the file you have downloaded in fact is the data set you wanted to get, i.e., is it an authentic copy? You may be able to verify this with various checksums, both on the level of the file (e.g. a hashcode of the file, as commonly provided for downloads) and on the level of observations in some cases. In this case, you are concerned with **authenticity** - you want to verify that the data set is what it purports to be.

Authenticity does not guarantee you, however, that the observations in the data set are any good. A good observation of weather data is one that you can rely on to accurately represent how the weather actually was at the temporal and spatial locations covered by the data. In other words, when you want the data set to be able to stand in for the facts it represents, you are concerned with **reliability.** In other words, reliability is very much about the relationship of the data to whatever it represents. If the data set is a compilation of social media posts, then reliability will relate to the question whether these contributions were really posted, etc.

**Integrity** on the other hand refers to questions of tampering, errors, etc. For example, a dataset that lacks integrity is one for which we can not assert that it contains *all* the items it originally contained, or that none of the items have been altered, falsified, or faked.

Consider a textbook case for records and archives for the difference between the three. A *passport* is a document that comes with very special features to prove that it can *stand in for the fact* that you are a citizen of the issuing country. Its **integrity** refers to the question whether it has been tampered with - has the photo been peeled off, have pages been removed or added? etc. The passport comes with features to prevent and check integrity. Its **authenticity** refers to the fact that it is indeed a passport of that country and that it indeed asserts the facts it states. Most of its special features are designed to make it easy to verify that (cf. banknotes). But imagine: a government could issue a perfectly authentic passport for a person who doesn't exist. That would be authentic, but it would not be reliable. The **reliability** rests on the relationship to the person it represents. We trust an authentic passport to be reliable because we trust the processes that governments have instituted and honed over the centuries to ensure that passports are only *issued to* authenticated citizens. But border control will use a machine readable passport to look up and

compare the information shown with the information stored in a database. When they do that, they verify reliability. For a deep dive into the archival perspective on what makes records authentic and reliable, see [32,36].

Finally, **representativeness** is related to reliability but its perspective is much more narrowly focused on the question whether a data set accurately *represents* the overall set of observations or entities that it claims to be a sample of. For instance, for a data set of social media posts, the question will arise if it's representative of all platforms, all users, all topics, all media types, or various combinations of dimensions. All the statistical concepts around sampling apply as usual. Other data sets are not sampled out of an identified population but claim to stand for a general category so that representativeness is evaluated analytically, and so on.

### How to interpret findability, accessibility, interoperability, and reusability (FAIR)

Note that this group of criteria are a direct representation of the widely used FAIR principles [138] for research data sets, adopted and adapted for machine learning. We provide a simple checklist to assess whether the documentation of the dataset discusses the application of FAIR principles. This checklist is derived from the following tools and resources:

- Minglu Wang and Dany Savard. 2023. The FAIR Principles and Research Data Management. (September 2023). https://doi.org/10.5206/EXFO3999
- FAIR data maturity model
- https://zenodo.org/records/5111307#.Yj3Vi5rMI-Q
- https://ardc.edu.au/resource/fair-data-self-assessment-tool/
- https://fairaware.dans.knaw.nl/

1. Findable
   a. A globally unique (cannot be reused by someone else) and persistent (valid over time) ID (like DOI) is assigned to the data.
   b. The dataset is described by metadata (PID, license, description, provenance, etc.). Further guidelines and definitions of provenance can be found from the DCMI and our glossary.
   c. The metadata specifies the identifier.
   d. The metadata and data is stored in a searchable repository.
2. Accessible
   a. The identifier navigates to the metadata and data.
   b. Retrieval of the data is specified by a standard communications protocol (i.e., all information and tools that are required are communicated to access the content of the dataset) which is open and free to access.
   c. The communications protocol specifies the authentication and authorization procedure, if needed (i.e., if the dataset is not open and free-to-access, the protocol specifies how access would be granted).
   d. The metadata record is available even if the data is not.
3. Interoperable
   a. Metadata and data are *in principle* readable by humans and machines (i.e., has a structured format, open standard).
   b. Metadata and data use controlled vocabularies (standardized and universal terms for indexing and information retrieval). Metadata standards can be found in the RDA Metadata Standards Catalog (https://rdamsc.bath.ac.uk/).
   c. Metadata and data is linked to other metadata and data using qualified references (i.e., relationship to the resource is specified).
4. Reusable
   a. Metadata and data are well-described as per domain-relevant standards, have detailed provenance (where did the data come from, who collected it, when, etc.), and clear and accessible license and usage information.

### Guiding Principles

We specify the following principles as "rules of thumb" to guide the evaluation of datasets:

1. Evaluate explicit documentation

Evaluations should be made on the basis of documentation provided by the dataset creators, rather than performing evaluations ourselves.

2. Provide traceable comments

The comments provided in the rubric to support the grade for each element should make recoverable the basis for the evaluation.

3. Minimum is easy, excellence is hard.

The evaluations for the minimum standard are meant to be *generous*. The evaluation should consider any amount of documentation as a sufficient indicator of reflection for that element. Therefore, meeting the minimum standard should be relatively easy. On the other hand, the standard of excellence criteria advocates for a high level of criticality, which is significantly

harder to attain (compared to the minimum standard). The evaluations should therefore only grant a 'Full' if all criteria are satisfied.

4. Don't make excuses.

If there is no documentation provided to evaluate an element, then don't make excuses for the dataset creators and evaluate it yourself or think of it as unnecessary. If you truly feel the element does not apply for that dataset, then that means it's feedback for the rubric and that the element needs further work so it applies to all types of datasets.

**Reflections & recommendations**

In addition to the instructions on the process of using the rubric to evaluate datasets, the following recommendations are provided based on common reflections, challenges, and questions:

1. Completing an evaluation using the rubric requires iteration. A single pass through the rubric is often insufficient especially for datasets that include various sources of documentation. The first iteration should be a step-by-step completion of each element in the rubric by looking for relevant information, keywords in the research paper or other dataset documentation. However, in doing so, sections of the documentation may be missed. It is therefore suggested to first evaluate the dataset by applying the rubric sequentially and then reviewing all the dataset documentation sequentially. The final step should be iterating as needed and zooming out.
2. The evaluation of elements will be interconnected. In example 1, there is an authenticity issue where the summaries may or may not accurately represent the actual court case. This is linked to the lack of reflection on the impact of data practices as well as a missing discussion on how disagreements between annotators are resolved. Since all of these elements are linked together, there can be a note to refer to the comment for another element.
3. If a context document is provided, it must be used to evaluate the elements. Although, the document will only provide information to fill in gaps rather than be sufficient to completely evaluate any element.
4. None of the elements should receive an N/A comment or grade.
5. The standard of excellence criteria should only be evaluated if the minimum standard criteria passes.
6. A failure for any element should be not provided based on the quality of the dataset but rather the documentation and reflection on the process of developing the dataset. For example, if the documentation acknowledges that the sample is not representative and can therefore introduce a bias- this is not considered a 'Fail'.
7. It is important to not evaluate the technical details provided but only evaluate the documentation. This means that evaluators should refrain from inferring the thought process or intention of the dataset creators based on their technical understanding of why the creators would develop their dataset in one way versus another. It is key to rely on the explicit documentation only. This is important because the rubric assesses critical reflection around the dataset process not the quality of the dataset developed.

**FAQ**

1. Is there a difference between labeling and annotation?

Please refer to the glossary for definitions differentiating the two terms. The rubric doesn't require evaluation of the "labeling" process if the dataset does not have labels.

2. How to evaluate consistency and timeliness for suitability?

Data quality is often defined as fitness for purpose and is multi-dimensional, meaning that it's measured through more than one data quality dimension such as accuracy, completeness, etc. Suitability, in the rubric, evaluates whether dataset creators ensure that their dataset's quality meets the purpose defined. For example, a dataset of math problems may not require timely data but may require consistent data (i.e., data presented in the same format). For standard of excellence, multiple data quality dimensions will apply for evaluation but potentially not all.

3. Is representativeness applicable to synthetic data?

Representativeness is still applicable to synthetic datasets because synthetic data is still representative of reality. However, this is a *conceptual* representativeness rather than a *statistical* one.

4. Why does the evaluation criteria for authenticity discuss data processing specifically?

Data processing alters the authenticity of a digital object. Authenticity is dependent on the bits of information in a file. For example, if you download a dataset with a hash code and make copies of it, all copies will have the same hash code. However, if you perform data processing (which changes the bits), the hash code will no longer be the same. In the rubric, for the minimum standard, you evaluate whether the dataset creators validate and verify the authenticity of the data they are collecting. Whereas for standard of excellence, you evaluate whether they have processes to ensure people that reuse their dataset are able to claim authenticity (i.e., maintaining the chain of authenticity).

5. For the data quality elements, are we evaluating that the dataset is suitable, authentic, has integrity, is representative, and is reliable OR that the dataset creators discuss their processes for ensuring these? If there is no mention of these qualities specifically, how do we evaluate them?

For data quality elements, you are evaluating whether the dataset creators discussed their processes for ensuring that their dataset is suitable, authentic, reliable, has integrity, and the extent to which it is representative (and why if it is not). Remember

the guiding principle- "evaluate explicit documentation". We have added another guiding principle- "don't make excuses". If no documentation is provided for these data quality elements, then don't make excuses for the dataset creators and evaluate it yourself or think of it as unnecessary. If you truly feel the element does not apply for that dataset, then that means it's feedback for the rubric and that the element needs further work so it applies to all types of datasets.

6.    Does hosting a dataset on huggingface make it 'findable'?
It depends, if it's hosted on huggingface but does not have a persistent identifier like a DOI, then it is not findable. See next question.

7.    Why are URLs not acceptable for findability?
URLs are not considered "findable" because of the high likelihood of link rot (that the link over time will no longer be available). There are studies that show that academic papers are highly perceptible to link rot, eg: see [74]. Instead, we want persistent identifiers like DOIs to make sure the dataset is findable in the future.

8.    What is the difference between findability and accessibility?
Findability is about a dataset being easily located. For example, if a publication provides a zenodo link to a dataset, that would make it findable (zenodo assigns a DOI to everything it publishes). So here we're looking for a dataset being easily located, indexed, catalogued, etc.

Accessibility is about whether a dataset can be opened and used and read. For example, is it in a format you can read, can you download it (i.e., is it retrievable), is the access blocked off via password-protection, are there access and authorization protocols?

A dataset would then be findable if there was a link pointing to it but not accessible if you couldn't open it because you didn't have the password for it and there was no documentation of an access protocol. On the other hand, if a dataset was open-access (eg, through github) but didn't have a persistent identifier (eg DOI) and wasn't indexed in a repository like zenodo then it would be accessible but not findable. Since accessibility rests on *accessing* the content, a URL alone is not enough to make it accessible either. So even if the dataset is available through github there must be other documentation that provides any further information needed to access the content and metadata.

9.    Can you provide further clarification for evaluating interoperability (especially standard of excellence)?
For the minimum standard, the documentation must explain how the dataset can be integrated with other data and workflows. An example of that is that the data can be exported to popular, standard formats. For the standard of excellence, the data and metadata must use controlled vocabularies and link to other resources with qualified references. For example, metadata can be created using controlled vocabularies like the W3C's Data Catalog Vocabulaire (DCAT) model which defines terms like dataset vs data service, catalog (as a subclass of dataset), and so on. Please see this blurb from FAIR about qualified references:
> "A qualified reference is a cross-reference that explains its intent. For example, *X is regulator of Y* is a much more qualified reference than *X is associated with Y*, or *X see also Y*. The goal therefore is to create as many meaningful links as possible between (meta)data resources to enrich the contextual knowledge about the data, balanced against the time/energy involved in making a good data model. To be more concrete, you should specify if one dataset builds on another data set, if additional datasets are needed to complete the data, or if complementary information is stored in a different dataset. In particular, the scientific links between the datasets need to be described. Furthermore, all datasets need to be properly cited (i.e., including their globally unique and persistent identifiers)." [44].

Zenodo also has a webpage that describes how it fulfills the FAIR principles for its datasets [143].

**Rubric**

The rubric can be found in Appendix A.1.

**Rubric Worksheet**

The rubric worksheet can be found in Appendix A.2.

**Sample Evaluations**

Please note that the sample evaluations were performed using the version of the rubric at the time of evaluating datasets from round 3. Note also that the description column and cited references are deleted below for space, see full rubric with references.

**Example 1**

Paper: FS-Mol: A Few-Shot Learning Dataset of Molecules [130]

| | CURATORIAL ELEMENT | DOCUMENTATION LEVEL | | | |
|---|---|---|---|---|---|
| | | Criteria to meet minimum standard | PASS/ FAIL | Criteria to meet standard of excellence | Full/ Partial/ None |
| | | | SCOPE | | |
| 1 | Context, purpose, motivation | Pass | Paper introduction discusses the problem domain and why a new dataset is needed; see 'related work' in paper and appendix B in supplementary material ('related work details') for comparison to existing datasets. | Full | Section 7 of paper discusses how dataset can be used outside of its original context ("it is now possible to evaluate… we note that transfer of results to realistic projects is not guaranteed to be successful…") |
| 2 | Requirements | Pass | Section 2 of paper (especially " 2.2 Desired Attributes of a QSAR Few-Shot Dataset and Benchmark") explicitly derives design requirements to create the dataset. | Partial | No explicit discussion of intrinsic biases introduced by problem formulation; other approaches to formulating the problem are discussed in 'related work' section of paper (discussing other datasets and their features) |
| | | | ETHICALITY AND REFLEXIVITY | | |
| 3 | Ethicality | Pass | No discussion of consent (no human data); pg 9 'societal impacts' section discusses benefits of creating the dataset. | Fail | No additional discussion of ethical consideration throughout the paper or supplementary documentation. |
| 4 | Domain knowledge & data practices | Pass | On pg 2 of papers, authors state aim to "demonstrate the utility of few-shot learning methods in an important domain, namely QSAR, which does not provide an obvious generic pretraining corpus (such as in NLP or computer vision). The proposed dataset is specifically designed to replicate the challenges of machine learning in the very low data regime of drug-discovery projects" (focus on drug-discovery domain) | Partial | README in GitHub repo discusses activities to be undertaken to re-use the dataset "Hence, in order to be able to run MAT, one has to clone our repository via…" – not directly discussing any domain knowledge needed. |
| 5 | Context awareness | Fail | Research goals are described but not positioned relative to researchers' intellectual/political believes; researcher positions not disclosed/no positionality statement included. | None | Failed minimum criteria. |
| 6 | Environmental footprint | Fail | No assessment of environmental footprint | None | Failed minimum criteria |
| | | | DATA PIPELINE | | |
| 7 | Data collection | Pass | ExtractDataset.ipynb from GitHub repo describes how data were gathered by querying ChEMBL; section 3 of paper explains data acquisition process in detail ("the reason why we remove large assays is…") | Partial | Section B of supplementary material describes other few-shot learning and molecular property datasets (e.g. why they used ChEMBL instead of other sources); no explicit discussion of criteria for source selection, why criteria were chosen, or how other sources were validated against criteria. |
| 8 | Data processing | Pass | ExtractDataset.ipynb from GitHub repo describes how data were cleaned and split into test vs validation assays. | Full | Section 3 of paper describes decisions behind data processing (e.g. "In this way, our proposed meta-testing tasks closely mimic the new-lead optimization problem, where a completely unseen task is presented for adaptation.") |
| 9 | Data annotation | Pass | "Binary Classification Task" section of paper discusses some annotation activity | None | No discussion of robustness of annotations. |

| | CURATORIAL ELEMENT | DOCUMENTATION LEVEL | | | |
|---|---|---|---|---|---|
| | | Criteria to meet minimum standard | PASS/ FAIL | Criteria to meet standard of excellence | Full/ Partial/ None |
| DATA QUALITY | | | | | |
| 10 | Suitability | Pass | Section 6 and first paragraph of section 7 describe and demonstrate dataset appropriateness for purpose. | Partial | Documentation does not explicitly discuss accuracy/completeness/timeliness of the chosen dataset, but Section 6 of the paper demonstrates the utility of the dataset for its intended purpose by providing "a set of results for all three categories of few-shot learning, with representative methods of the use of this dataset in each". |
| 11 | Representativeness | Pass | Section 3 on pg 3 of main paper describes how the 'sample' of the dataset is taken from the overall population (the ChEMBL database); also on pg 9 "the few-shot baselines we provide checkpoints and results for are only a representative set, rather than a complete survey of the current state of the field" | None | No explicit discussion of biases. |
| 12 | Authenticity | Pass | No explicit discussion of authenticity but extractdataset.ipynb does discuss how initial raw data were obtained (e.g. describes process by which database was queried) | Partial | No explicit discussion of future authenticity/preservation processes, but does discuss in section A of supp material how dataset documentation facilitates re-use more generally. |
| 13 | Reliability | Pass | Section 5 of paper discussing benchmarking procedures (i.e. making sure that the dataset is useful for what it's supposed to be useful for) | Partial | No explicit discussion of reliability management in the context of future re-use; section A of supplementary material discusses how the dataset documentation facilitates re-use. |
| 14 | Integrity | Fail | No discussion of dataset integrity or preservation processes (section H of supplementary document does not actually discuss a maintenance plan or means of maintaining accuracy/consistency over time). | None | Failed minimum criteria. |
| 15 | Structured documentation | Fail | No standardized context document | None | Failed minimum criteria |
| DATA MANAGEMENT | | | | | |
| 16 | Findability | Fail | No persistent identifier provided. | None | Failed minimum criteria |
| 17 | Accessibility | Pass | Section F of supplementary material describes computational resources used; GitHub README states the tools and steps required to access data content. | Partial | GitHub repo includes a code of conduct document, as well as protocols for contributing and for security reporting. |
| 18 | Interoperability | Pass | README in GitHub repo describes how to use the dataset with "three key few-shot learning methods"; dataset.ipynb describes the machine/human readable metadata. | Full | Dataset.ipynb describes the controlled vocabularies for specific dataclasses (e.g. task_name as a string describing the task each point is taken from) |
| 19 | Reusability | Fail | From data contents of GitHub repo it does not appear that data or metadata contain provenance information about where the dataset came from/when/who collected it; license is included in the GitHub repo. | None | Failed minimum criteria. |

**Example 2**

Paper: American Stories: A Large-Scale Structured Text Dataset of Historical U.S. Newspapers [26]

| | CURATORIAL ELEMENT | DOCUMENTATION LEVEL | | | |
|---|---|---|---|---|---|
| | | Criteria to meet minimum standard | PASS/ FAIL | Criteria to meet standard of excellence | Full/ Partial/ None |
| | | | SCOPE | | |
| 1 | Context, purpose, motivation | Pass | Paper Introduction and section 6 (Applications) discusses the problems and relevance, and 'Related Literature' (section 2) discusses other similar datasets. | Full | 'Applications' section on pg 6 of supplementary material discusses "multiple applications that can be facilitated by the American Stories dataset" |
| 2 | Requirements | Pass | Paper Introduction (pg 2, "To address these limitations, we develop…") introduces certain requirements. | Partial | On pg 3 of paper ,documentation reflects on the bias potentially introduced by scanning illegible newspapers; other approaches are discussed in Section 2 on Related Literature (but not specifically other approaches the authors considered) |
| | | | ETHICALITY AND REFLEXIVITY | | |
| 3 | Ethicality | Pass | Some harms (e.g. offensive language) are discussed in Section 7: Conclusion. Consent is discussed in datasheet (pg 14 of supplementary material) | Partial | Some additional discussion of copyrights/accessibility on pg 3 of paper |
| 4 | Domain knowledge & data practices | Pass | Pg. 23 of paper (the datasheet) addresses the professors, research assistants, and students involved in data collection | Partial | Datasheet states "There are a large number of potential uses in the social sciences, digital humanities, and deep learning research" |
| 5 | Context awareness | Pass | No positionality statement but several mentions throughout the datasheet showing awareness of social context ("This dataset contains unfiltered content composed by newspaper editors, columnists, and other sources. It reflects their biases and any factual errors that they made."), and section 7 of the paper reflects on the historicity of dataset contents | Partial | Section 3 of paper touches on assumptions going into methodological choices (e.g. on pg 3, "We do not OCR ads because…") |
| 6 | Environmental footprint | Fail | No environmental assessment. | None | Failed minimum criteria |
| | | | DATA PIPELINE | | |
| 7 | Data collection | Pass | Described in 'Composition' (pg 11) and 'Collection Process' (pg 13) sections of datasheet in supplementary material | Partial | We have a lot of information about how the data were collected, but I still don't see where in the documentation it specifies the criteria they used to select data sources or how data sources were validated against these criteria (e.g. why the library of congress dataset?). |
| 8 | Data processing | Pass | Pre-processing section of datasheet (pg 14 of supplementary material) describes process of cleaning and wrangling data | Full | Sections 3, 4, and 5 of main paper discuss the implications of processing decisions (e.g. on computing cost and efficiency) |
| 9 | Data annotation | Pass | Student annotation is discussed ins Section 5 'Pipeline Evaluation' of main paper | Full | Student annotations were used as 'ground truth' for model training; see pg 5 of supplementary material |
| | | | DATA QUALITY | | |
| 10 | Suitability | Pass | Section 5 of paper evaluates the pipeline for accuracy, legibility, and comparison to other OCR engines | Full | See explanation for minimum criteria |
| 11 | Representativeness | Pass | Sampling approach discussed in datasheet (pg 13 of supplementary material) – it includes everything in the Chronicling American scan collection. | Full | Section 3 of paper discusses how illegible papers and their inclusion/exclusion in the dataset could bias results. |
| 12 | Authenticity | Pass | Pipeline for generating data is included in the Github repo (https://github.com/dell-research-harvard/AmericanStories?tab=readme-ov-file); no explicit discussion of authenticity | None | No explicit discussion of authenticity in future re-use. |

| | CURATORIAL ELEMENT | DOCUMENTATION LEVEL | | | |
|---|---|---|---|---|---|
| | | Criteria to meet minimum standard | PASS/ FAIL | Criteria to meet standard of excellence | Full/ Partial/ None |
| 13 | Reliability | Pass | Section 5 of paper (Pipeline Evaluation) describes verification and validation processes used to ensure reliability. | Full | Maintenance section of datasheet discusses how errors will be corrected in future (and uploaded to HuggingFace) |
| 14 | Integrity | Pass | Documentation does not explicitly discuss integrity but datasheet does emphasize that "material is complete and unaltered" | Full | Maintenance section of datasheet describes preservation processes in place (e.g. old versions still accessible via HuggingFace) |
| 15 | Structured documentation | Pass | Paper and supplementary material include a datasheet (Gebru et al) | Full | All mandatory components of datasheet are answered. |
| | DATA MANAGEMENT | | | | |
| 16 | Findability | Pass | DOI available on HuggingFace page (10.57967/hf/0757) | Full | Data and metadata stored in searchable repo (HuggingFace) |
| 17 | Accessibility | Pass | Steps for accessing data listed on HuggingFace page data card and described in 'Distribution' section of datasheet (pg 15 of supplementary material | Full | Communications protocol described in 'Maintenance' section of datasheet (supp material pg 16) |
| 18 | Interoperability | Pass | Pg 4 of paper describes readable formats of metadata and data ("The raw files are in a json format, and the Hugging Face repo comes with a setup script that easily allows people to download both raw and parsed data to facilitate language modeling and computational social science applications."; lots of metadata info included on HuggingFace page | Full | See HuggingFace page for controlled metadata vocabularies |
| 19 | Reusability | Pass | Some provenance information included in metadata (e.g. where it came from, associated newspaper, but not who collected it/when) | Partial | Pg 16 of supplementary material (datasheet) states "The dataset is distributed under a Creative Commons CC-BY license. The terms of this license can533 be viewed at https://creativecommons.org/licenses/by/2.0/" |

**Glossary**

| Term | Definition | Discussion/Example | Sources |
|------|-----------|-------------------|---------|
| Context document | "Interventions designed to accompany a dataset or ML model, allowing builders to communicate with users". | Context documents are standardized documentation formats that convey information about the dataset, types of context documents include datasheets, nutrition labels, etc. | [19] |
| Data annotation | Although data annotation and labelling are often used interchangeably in ML, labelling is a subset of annotation. (See labelling)<br><br>Data annotation refers to the process of adding information to a dataset to provide more context. For example, adding metadata. | Annotation can include metadata about the units of measurement. | |
| Data practices | "What and how data are collected, managed, used, interpreted, released, reused, deposited, curated, and so on…" | Data practices are the decisions made in the collecting, interpretation, etc. of data. | [18] |
| Extrinsic bias | Extrinsic bias refers to bias that exists within the dataset which are reflections of social, historical biases. | "Extrinsic bias is concerned with a view of a biased dataset "from the outside." The argument is that an already-biased dataset can cause even innocent software to produce a biased outcome - and may look like people saying things such as "the data made me do it." … If we fail to remember that a dataset is biased, then we may treat it as "fair" or "representative," harming people who have been excluded from it." | [105] |
| Informed consent | Informed consent is a standard ethical principle of research with human subjects that rests on the commitment that participants<br>• Are fully informed<br>• Decide voluntarily<br>• Before research is conducted.<br>Its application in online environments is complicated by the shift in technology and methods (see [36]), but the principle remains important. | In conventional human subject studies such as interviews, an IRB reviews ethics protocols and evaluates if the research is compliant with principles such as the proportionality principle.<br>In social media research, things get complicated. In some situations, implied consent (see [64]) can be present but must be justified. In the case of LLMS, widespread data collection without consent has prompted massive ethical and legal concerns. | A discussion of Twitter research ethics [36] is a good start. |
| Intrinsic bias | "The ways in which we change the data "from the inside" of data science work-processes while we are preparing the data for modeling." Intrinsic bias is the bias data workers introduce to the dataset. | "Through practices of data wrangling, curation, and feature-engineering, humans make a series of decisions about how to treat their data, and those decisions may inadvertently introduce bias into the data." | [105] |
| Labelling | Labelling is a specific type of annotation that involves assigning a predefined category to a data item. | Labelling tweets on Twitter as 'human-generated' or 'bot-generated'. | |
| PID: persistent identifier | "Globally unique and persistent identifiers remove ambiguity in the meaning of your published data by assigning a unique identifier to every element of metadata and every concept/measurement in your dataset. [IDs] must be persistent. It takes time and money to keep web links active, so links tend to become invalid over time. Registry services guarantee resolvability of that link into the future, at least to some degree." | ORCID iDs are persistent identifiers for people. DOIs are persistent identifiers for journal articles, datasets, etc. | [43] |

| Population | Mathematical term used to describe a group of units sharing a common trait. | | |
|---|---|---|---|
| Positionality statement | "Researcher/Practitioner Self-Disclosure: Practice should involve a disclosure of the researcher's position in the world, her or his goals, as well as the researcher's position in her or his intellectual and, to an appropriate extent, political beliefs" | See [85] for examples. | [8] |
| Proportionality principle | In ethics, it is understood that actions have positive and negative effects simultaneously. This is called *double effect*. "Applications of double effect always presuppose that some kind of proportionality condition has been satisfied. Traditional formulations of the proportionality condition require that the value of promoting the good end outweigh the disvalue of the harmful side effect." | In medicine, a surgeon may cause harm to a patient's skin (negative) in order to save their heart (positive). It would not be permissible for a surgeon to open up someone's chest just to get a better look or take a selfie with it because that would violate the proportionality principle. | [97] |
| Provenance | Provenance information provides a trail of history about how the data originated, how it's changed, who was involved, and more. | See the following blurb from the FAIR principles… "For others to reuse your data, they should know where the data came from ... who to cite and/or how you wish to be acknowledged. Include a description of the workflow that led to your data: Who generated or collected it? How has it been processed? Has it been published before? Does it contain data from someone else that you may have transformed or completed?" | [45] |
| Reflexivity | "Questions of reflexivity ask us to consider who we should listen to and why, how to place actors' ideas in a larger field of power, questions about our own relationship to actors' theories of the world. Reflexivity asks us to approach our work with epistemological unease because we are always at risk of reproducing categories that reify power." | | [99,132] |

**Further Readings**

The following readings 1) showcase how data curation is discussed in data science and machine learning studies, 2) contain context for relevant data curation terms, concepts, and frameworks, and 3) provide important terminology for ML benchmarks. Readings are listed as required and suggested.

**Data Curation in Data Science**

A vast amount of literature points to the datasets used for training machine learning models to be the source for introducing bias in model results leading to a call for increased documentation of datasets used in ML. Emerging research has proposed context documents – "interventions designed to accompany a dataset or ML model, allowing builders to communicate with users". The following are types of relevant context documents.

Required:
1. Timnit Gebru, Jamie Morgenstern, Briana Vecchione, Jennifer Wortman Vaughan, Hanna Wallach, Hal Daumé III, and Kate Crawford. 2021. Datasheets for datasets. *Commun. ACM* 64, 12 (November 2021), 86–92. https://doi.org/10.1145/3458723

Datasheets are one of the most popular methods of documenting the process of developing datasets as well as providing a dataset description. This paper is a good introduction to how dataset documentation is evaluated [40].

Suggested:
2. Michael A. Madaio, Luke Stark, Jennifer Wortman Vaughan, and Hanna Wallach. 2020. Co-Designing Checklists to Understand Organizational Challenges and Opportunities around Fairness in AI. In *Proceedings of the 2020 CHI Conference on Human Factors in Computing Systems*, April 21, 2020, Honolulu HI USA. ACM, Honolulu HI USA, 1–14. https://doi.org/10.1145/3313831.3376445

Madiao et al. developed a resource - checklist for AI fairness - based on findings of current practitioners processes, needs, and requirements for developing fair AI models [92].

3. Emily M. Bender and Batya Friedman. 2018. Data Statements for Natural Language Processing: Toward Mitigating System Bias and Enabling Better Science. *Transactions of the Association for Computational Linguistics* 6, (2018), 587–604. https://doi.org/10.1162/tacl_a_00041

Bender and Friedman develop 'data statements'- a resource for NLP training datasets to be documented in order to mitigate bias and exclusion [12].

Topics like dataset documentation in ML are often discussed as a part of data practices, data work, or dataset development. The following studies talk about stages of dataset development processes, how data scientists or data workers approach their data work, and the importance and impact of decisions made during the dataset development.

Required:

1. Morgan Klaus Scheuerman, Alex Hanna, and Emily Denton. 2021. Do Datasets Have Politics? Disciplinary Values in Computer Vision Dataset Development. *Proc. ACM Hum.-Comput. Interact.* 5, CSCW2 (October 2021), 1–37. https://doi.org/10.1145/3476058

This paper discusses how documentation captures underlying values of data practices in machine learning (specifically computer vision tasks) [125]. Specifically, publications are analyzed to understand the documentation and communication of datasets. The findings showcase the practices that are silenced (such as data work, context, positionality, and care) over those that are (wrongly) embraced such as model work, universality, and so on. This reading help reflect on and understand how intrinsic bias can be introduced within datasets.

Suggested:

2. Nithya Sambasivan, Shivani Kapania, Hannah Highfill, Diana Akrong, Praveen Paritosh, and Lora M Aroyo. 2021. "Everyone wants to do the model work, not the data work": Data Cascades in High-Stakes AI. In *Proceedings of the 2021 CHI Conference on Human Factors in Computing Systems*, May 06, 2021, Yokohama Japan. ACM, Yokohama Japan, 1–15. https://doi.org/10.1145/3411764.3445518

Through interviews with AI practitioners, Sambasivan et al. find that poor data practices in high-stakes AI domains (i.e., practices that do not prioritize data quality) lead to data cascades which are negative impacts of data issues [124].

3. Milagros Miceli, Julian Posada, and Tianling Yang. 2022. Studying Up Machine Learning Data: Why Talk About Bias When We Mean Power? *Proc. ACM Hum.-Comput. Interact.* 6, GROUP (January 2022), 1–14. https://doi.org/10.1145/3492853

Miceli et al. discuss that while we often recognize that there is bias in the datasets and their processes used for ML models, it is often ignored that this bias is a result of power inequities [99]. The authors analyze data bias, data work, and data documentation from a "power-aware" framing as compared to a "bias-oriented" one. This paper provides an interesting shift in perspective which further illuminates the importance of reflexivity in data work.

4. Michael Muller and Angelika Strohmayer. 2022. Forgetting Practices in the Data Sciences. In *CHI Conference on Human Factors in Computing Systems*, April 29, 2022, New Orleans LA USA. ACM, New Orleans LA USA, 1–19. https://doi.org/10.1145/3491102.3517644

This paper studies how data processing leads to different types of forgetting and where and how each type of forgetting occurs in the machine learning stack [105]. Forgetting is conceptualized as the practice that occurs when choices are made about what data is kept, what it represents and so forth (therefore by designing a dataset in a given way, we *remember* only its current state, and *forget* the decisions, the erased data, etc.). This is a great paper for a deep dive into the various types of design decisions that impact the eventual dataset.

The previous studies discuss aspects of data curation as dataset development. However, some ML studies have started discussing the importance of data curation by referencing archival studies and digital curation directly. These are included below:

Required:

1. Susan Leavy, Eugenia Siapera, and Barry O'Sullivan. 2021. Ethical Data Curation for AI: An Approach based on Feminist Epistemology and Critical Theories of Race. In *Proc. of 2021 AAAI/ACM Conf. on AI, Ethics, and Society*, July 21, 2021, Virtual Event USA. ACM, Virtual Event USA, 695–703. Retrieved November 11, 2022 from https://dl.acm.org/doi/10.1145/3461702.3462598

This study discusses principles for ethical data curation based on race critical race theory and data feminism to improve the reflection of power, bias, and values in data processes and thereby improve transparency and accountability of AI systems [80].

Suggested:

2. Emily M. Bender, Timnit Gebru, Angelina McMillan-Major, and Shmargaret Shmitchell. 2021. On the Dangers of Stochastic Parrots: Can Language Models Be Too Big? 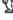. In *Proceedings of the 2021 ACM Conference on Fairness, Accountability, and Transparency* (*FAccT '21*), March 01, 2021, New York, NY, USA. Association for Computing Machinery, New York, NY, USA, 610–623. https://doi.org/10.1145/3442188.3445922

This paper discusses the potential risks of language models (and by extension other ML/AI systems) [13]. The authors recommend a shift towards careful, reflective practices around datasets and model development along with a greater focus towards documentation.

3. Eun Seo Jo and Timnit Gebru. 2020. Lessons from archives: strategies for collecting sociocultural data in machine learning. In *Proceedings of the 2020 Conference on Fairness, Accountability, and Transparency*, January 27, 2020, Barcelona Spain. ACM, Barcelona Spain, 306–316. https://doi.org/10.1145/3351095.3372829

   This paper highlights that practices from archival studies have experience dealing with consent, power dynamics, transparency, and ethics and that these practices should be adopted into data collection and annotation practices in machine learning [67].

## Data Curation

Data curation involves "maintaining and adding value to digital research data for current and future use". The following studies introduce data/digital curation terminology and the data curation lifecycle model (parallel to ML model pipelines) with the aim to familiarize how the data curation field approaches data work.

Required:
1. Sarah Higgins. 2008. The DCC Curation Lifecycle Model. *International Journal of Digital Curation* 3, 1 (August 2008), 134–140. https://doi.org/10.2218/ijdc.v3i1.48

   The paper introduces the curation lifecycle model by emphasizing it as a lifecycle (as opposed to a linear process). Each stage of the model is briefly introduced [55].
2. Sarah Higgins. 2012. The lifecycle of data management. In *Managing Research Data* (1st ed.), Graham Pryor (ed.). Facet, 17–46. https://doi.org/10.29085/9781856048910.003

   This paper discusses each stage in depth including the tasks performed, how each stage leads to the next, and the expected outcomes [57].

Suggested:
3. Digital Curation Centre. Glossary. *Digital Curation Centre*. Retrieved January 21, 2024 from https://www.dcc.ac.uk/about/digital-curation/glossary

   This is a glossary of common digital curation terms - to be returned to as a resource, as needed [27].
4. Carole L Palmer, Nicholas M Weber, Trevor Muñoz, and Allen H Renear. Foundations of Data Curation: The Pedagogy and Practice of "Purposeful Work" with Research Data. 16.

   This is an introductory paper to the field of data curation and its place within archival studies, library studies, and computer science [110].

## Benchmarking in ML

Benchmarking is often not a well discussed topic in machine learning papers. The below list is compiled to introduce commonly used terms including: benchmark dataset, benchmark tasks, simulator, synthetic dataset, baseline method, benchmark suite, etc.

1. Matthew Stewart. 2023. The Olympics of AI: Benchmarking Machine Learning Systems. *Medium*. Retrieved January 21, 2024 from https://towardsdatascience.com/the-olympics-of-ai-benchmarking-machine-learning-systems-c4b2051fbd2b

   Explains terms benchmark, benchmark dataset, benchmark tasks, baseline method, and benchmark suite [94].
2. Ramona Leenings, Nils R. Winter, Udo Dannlowski, and Tim Hahn. 2022. Recommendations for machine learning benchmarks in neuroimaging. *NeuroImage* 257, (August 2022), 119298. https://doi.org/10.1016/j.neuroimage.2022.119298

   Explains benchmark term and concept [83].
3. Kim Martineau. 2021. What is synthetic data? *IBM Research Blog*. Retrieved January 21, 2024 from https://research.ibm.com/blog/what-is-synthetic-data

   Explains term synthetic data [73].
4. Nataniel Ruiz. 2019. Learning to Simulate. *Medium*. Retrieved January 21, 2024 from https://towardsdatascience.com/learning-to-simulate-c53d8b393a56

   Explains term simulator [122].

## C        SUPPLEMENTARY INFORMATION ABOUT RUBRIC EVALUATIONS

## C.1        Positionality and Contributions

Our research team is composed of a combination of faculty members and graduate students at a Canadian university with a range of nationalities. A set of the authors (Eshta Bhardwaj, Tegan Maharaj, Christoph Becker) first conceptualized the rubric and toolkit prior to the evaluation phase. A different subset of the authors (Harshit Gujral, Siyi Wu, Ciara Zogheib) performed the evaluations using the rubric and toolkit. After starting the iterative evaluation process, all authors contributed to the

development and refinement of the framework and the resulting manuscript. Contributions are listed below using the Contributor Roles Taxonomy (http://credit.niso.org/) and additional details.

Eshta Bhardwaj (she/her) is a graduate student with expertise in data work in machine learning. She has studied how data curation concepts can be applied in dataset development in ML. Her contributions for this project include the conceptualization of the framework, aiding with project administration, developing the methodology of conducting the evaluations, analyzing and visualizing the results, and writing, editing, and reviewing all drafts of the published work.

Harshit Gujral (he/him) is a Ph.D. student in Computer Science at the University of Toronto, researching the health impacts of green energy. He has a background as a Data Scientist and an engineering degree in Information Technology. His contributions to this paper include development and iteration of the evaluations, conducting evaluations, analysis of structured documentation and its writing, writing – review & editing.

Siyi Wu (she/her) is a Ph.D. student in Computer Science at the University of Toronto, with a research focus on climate informatics and human-computer interactions. She has degrees in computer science and statistical science. Her contributions to this paper include development and iteration of evaluations, conducting evaluations, writing – review & editing.

Ciara Zogheib (she/her) is a graduate student at the University of Toronto's Faculty of Information, with a research focus on the information practices of data work. She has professional experience as a data scientist in government settings, including work developing ML governance strategies. Her contributions to this paper include development and iteration of the evaluations, conducting evaluations, writing – review & editing.

Tegan Maharaj is an Assistant Professor in the Faculty of Information at the University of Toronto and an affiliate of the Vector Institute and Schwartz-Reisman Institute for Technology and Society. Her contributions to this paper include funding, methodology, comments on iterations of the rubric, writing – review & editing.

Christoph Becker (he/him) is an immigrant settler in Canada and professor of information with degrees in computer science and social sciences and long research experience in systematic assessment and benchmarking of computational and organizational digital curation processes. He directs the Digital Curation Institute. His contributions to this paper include conceptualization, resources, funding, methodology, supervision, validation, writing – original draft, writing – review & editing.

This combination of expertise shaped the evaluation process and findings in several ways. The absence of digital curation background knowledge in the evaluation team aids the suggestion that this framework bridges the disciplinary views and can be applied in a ML context. Multiple iterations of questions and answers led to refinements in how the rubric criteria were articulated, entries in the glossary, and the introduction of examples in the toolkit documentation. The difference in disciplines also contributed to the identification of challenges, especially the challenge of overlapping terms with non-shared meanings.

## C.2    Datasets

Table 2: NeurIPS Datasets used for Evaluation

| Dataset Number | Round | Publication Title | Reference | Publication Year |
|---|---|---|---|---|
| 1 | Training | Programming Puzzles | [126] | 2021 |
| 2 | Training | Open Bandit Dataset and Pipeline: Towards Realistic and Reproducible Off-Policy Evaluation | [123] | 2021 |
| 3 | Training | SciGen: a Dataset for Reasoning-Aware Text Generation from Scientific Tables | [103] | 2021 |
| 4 | Training | MOMA-LRG: Language-Refined Graphs for Multi-Object Multi-Actor Activity Parsing | [91] | 2022 |
| 5 | Training | CEDe: A collection of expert-curated datasets with atom-level entity annotations for Optical Chemical Structure Recognition | [60] | 2022 |
| 6 | Round 1 | LoveDA: A Remote Sensing Land-Cover Dataset for Domain Adaptive Semantic Segmentation | [136] | 2021 |
| 7 | Round 1 | RELLISUR: A Real Low-Light Image Super-Resolution Dataset | [1] | 2021 |
| 8 | Round 1 | Measuring Mathematical Problem Solving With the MATH Dataset | [53] | 2021 |
| 9 | Round 1 | DGraph: A Large-Scale Financial Dataset for Graph Anomaly Detection | [61] | 2022 |
| 10 | Round 1 | Change Event Dataset for Discovery from Spatio-temporal Remote Sensing Imagery | [93] | 2022 |

| 11 | Round 1 | CAESAR: An Embodied Simulator for Generating Multimodal Referring Expression Datasets | [65] | 2022 |
| 12 | Round 1 | GLOBEM Dataset: Multi-Year Datasets for Longitudinal Human Behavior Modeling Generalization | [139] | 2022 |
| 13 | Round 1 | ClimateSet: A Large-Scale Climate Model Dataset for Machine Learning | [70] | 2023 |
| 14 | Round 1 | BubbleML: A Multiphase Multiphysics Dataset and Benchmarks for Machine Learning | [50] | 2023 |
| 15 | Round 1 | DataComp: In search of the next generation of multimodal datasets | [38] | 2023 |
| 16 | Round 2 | The CPD Data Set: Personnel, Use of Force, and Complaints in the Chicago Police Department | [59] | 2021 |
| 17 | Round 2 | The Tufts fNIRS Mental Workload Dataset & Benchmark for Brain-Computer Interfaces that Generalize | [62] | 2021 |
| 18 | Round 2 | How Would The Viewer Feel? Estimating Wellbeing From Video Scenarios | [95] | 2022 |
| 19 | Round 2 | The RefinedWeb Dataset for Falcon LLM: Outperforming Curated Corpora with Web Data Only | [115] | 2023 |
| 20 | Round 2 | Stanford-ORB: A Real-World 3D Object Inverse Rendering Benchmark | [77] | 2023 |
| 21 | Round 3 | FS-Mol: A Few-Shot Learning Dataset of Molecules | [130] | 2021 |
| 22 | Round 3 | Evaluating Out-of-Distribution Performance on Document Image Classifiers | [79] | 2022 |
| 23 | Round 3 | Dungeons and Data: A Large-Scale NetHack Dataset | [48] | 2022 |
| 24 | Round 3 | VisAlign: Dataset for Measuring the Alignment between AI and Humans in Visual Perception | [82] | 2023 |
| 25 | Round 3 | American Stories: A Large-Scale Structured Text Dataset of Historical U.S. Newspapers | [26] | 2023 |

### C.3 Inter-rater reliability across elements

To compare reliability across elements, we focus on evaluations past the training round. Figure 3 shows perfect IRR for both criteria of 'environmental footprint' and the minimum criteria for 'context, purpose, motivation', 'data collection', and 'suitability'. Random agreement (i.e., ICC~0) can be seen for both criteria of 'requirements', and the minimum standard for 'data processing' and 'authenticity'. Systematic disagreement is seen for the minimum criteria of 'reliability'. The remaining elements vary in their ICC estimates. The scoping elements have the least agreement with a median ICC of 0.27 (poor agreement).

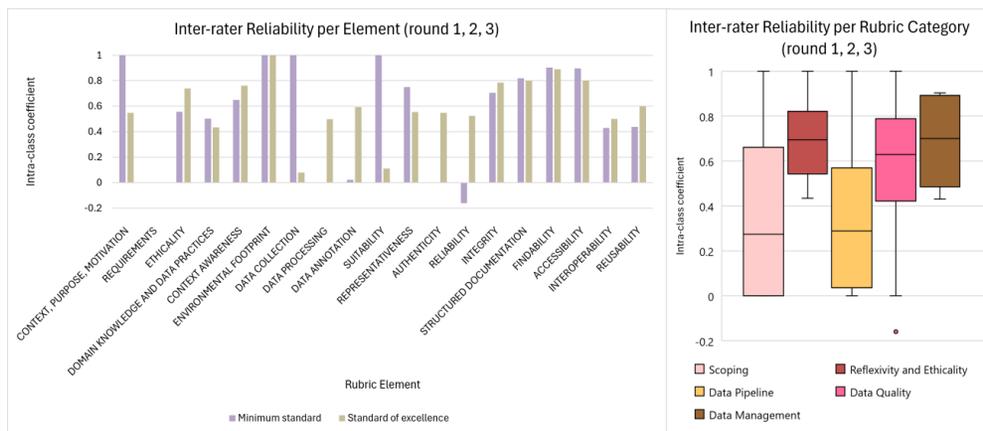

Fig. 3. IRR across elements and categories (round 1, 2, 3)

### C.4 Breakdown of inconsistencies by dataset

To assess the extent to which the rubric and toolkit improvements between the evaluations were impacting the consistency of the evaluations, we measured the number of "major" and "minor" disagreements in the evaluations for both the minimum standard and standard of excellence criteria for each round. The categorizations of disagreements were established as follows:

- Minor disagreement, standard of excellence (Minor, exc): instances where 1 of 3 reviewers disagree, e.g., Full, None, Full
- Major disagreement, minimum standard (Major, min): instances where 1 of 3 reviewers disagree, e.g., Pass, Pass, Fail
- Major disagreement, standard of excellence (Major, exc): instances all reviewers disagree, e.g., Full, Partial, None
- No disagreement, standard of excellence: instances where 1 of 3 reviewers gives a Partial evaluation and the other 2 reviewers agree, e.g., Partial, None, None

A summary of the average number of inconsistencies can be found in Figure 4. A further breakdown of the disagreements in evaluations for each of the 20 datasets is provided in Table 3. Figure 5 shows a consistent reduction in minor disagreements, some reduction in major disagreements for the minimum standard between training and round 1, and an increase again in round 2 and a significant reduction for round 3. The most reduction in disagreements is seen in major disagreements for the minimum standard. Overall, the percentage of all disagreements decreased from training (32%), round 1 (25%), round 2 (23%), to round 3 (7%).

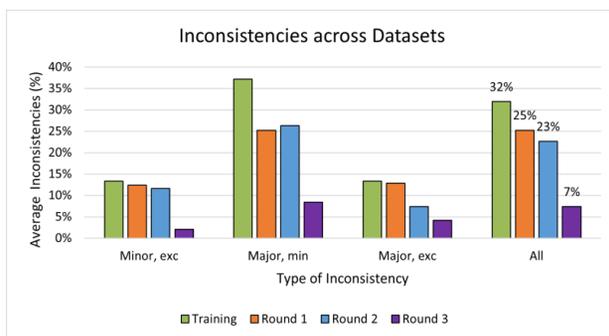

Fig. 4. Average percentage of inconsistencies across datasets

Table 3: Breakdown of inconsistencies by datasets

| | # | # Minor, exc | % Minor, exc | #Major, min | % Major, min | # Major, exc | % Major, exc | # Total | Average % |
|---|---|---|---|---|---|---|---|---|---|
| training (5 datasets, 21 elements across 2 levels = 42) | 1 | 3 | 14% | 11 | 52% | 2 | 10% | 16 | 38% |
| | 2 | 2 | 10% | 9 | 43% | 0 | 0% | 11 | 26% |
| | 3 | 0 | 0% | 1 | 5% | 5 | 24% | 6 | 14% |
| | 4 | 3 | 14% | 11 | 52% | 6 | 29% | 20 | 48% |
| | 5 | 6 | 29% | 7 | 33% | 1 | 5% | 14 | 33% |
| round 1 (10 datasets, 21 elements across 2 levels = 42) | 6 | 1 | 5% | 7 | 33% | 3 | 14% | 11 | 26% |
| | 7 | 2 | 10% | 3 | 14% | 3 | 14% | 8 | 19% |
| | 8 | 2 | 10% | 9 | 43% | 5 | 24% | 16 | 38% |
| | 9 | 3 | 14% | 10 | 48% | 1 | 5% | 14 | 33% |
| | 10 | 2 | 10% | 4 | 19% | 2 | 10% | 8 | 19% |
| | 11 | 3 | 14% | 4 | 19% | 2 | 10% | 9 | 21% |
| | 12 | 5 | 24% | 3 | 14% | 2 | 10% | 10 | 24% |
| | 13 | 2 | 10% | 6 | 29% | 3 | 14% | 11 | 26% |
| | 14 | 2 | 10% | 3 | 14% | 4 | 19% | 9 | 21% |
| | 15 | 4 | 19% | 4 | 19% | 2 | 10% | 10 | 24% |
| round 2 (5 datasets, 19 elements across 2 levels = 38) | 16 | 2 | 11% | 3 | 16% | 2 | 11% | 7 | 18% |
| | 17 | 2 | 11% | 5 | 26% | 0 | 0% | 7 | 18% |
| | 18 | 3 | 16% | 6 | 32% | 1 | 5% | 10 | 26% |
| | 19 | 2 | 11% | 3 | 16% | 2 | 11% | 7 | 18% |
| | 20 | 2 | 11% | 8 | 42% | 2 | 11% | 12 | 32% |
| round 3 (5 datasets, 19 elements across 2 levels = 38) | 21 | 0 | 0% | 2 | 11% | 0 | 0% | 2 | 5% |
| | 22 | 0 | 0% | 2 | 11% | 0 | 0% | 2 | 5% |
| | 23 | 0 | 0% | 1 | 5% | 1 | 5% | 2 | 5% |
| | 24 | 1 | 5% | 2 | 11% | 2 | 11% | 5 | 13% |

| | 25 | 1 | 5% | 1 | 5% | 1 | 5% | 3 | 8% |
|---|---|---|---|---|---|---|---|---|---|

## C.5    Breakdown of inconsistencies by element

We reviewed the inconsistencies across each element of the rubric to determine whether any specific elements were standing out as infeasible to adapt from data curation to ML or required further improvements to adapt. To measure this, we calculated the percentage of datasets with inconsistencies (i.e., the total number of datasets with any disagreement, where each dataset is counted only once even if there is more than 1 type of disagreement, divided by the total number of datasets in a given round), shown in Figure 5. The breakdown for each element across training, round 1, round 2, and round 3 can be found in Table 4. Figure 5 shows that 'context', 'purpose', 'motivation' and 'suitability' had no inconsistencies in all rounds. There were few elements for which the percentage of datasets with inconsistencies steadily reduced from training to round 3 ('ethicality', 'data processing', 'environmental footprint', 'findability', 'accessibility'). It was more common, though, for inconsistencies to change irregularly between rounds ('requirements', 'domain knowledge and data practices', 'context awareness', 'data collection', 'representativeness', 'authenticity', 'reliability', 'integrity', 'structured documentation', 'reusability') or for inconsistencies to increase ('data annotation', 'interoperability').

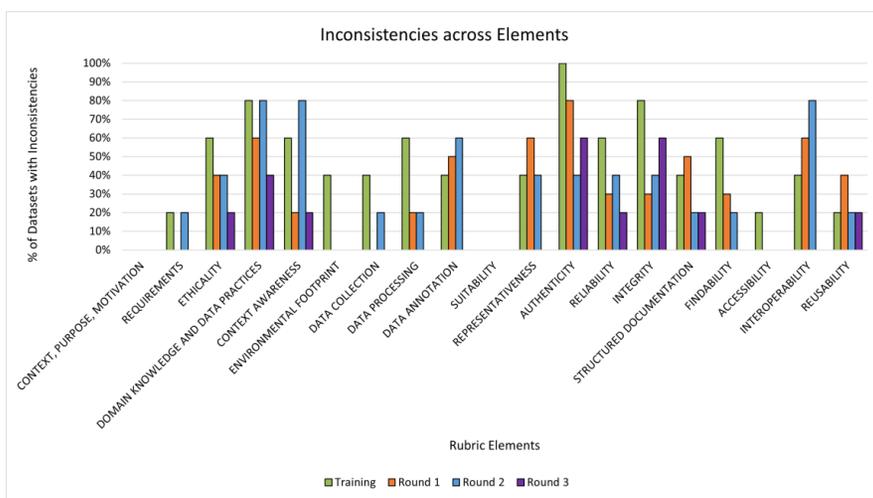

Fig. 5. Percentage of datasets with inconsistencies across elements

Table 4: Breakdown of inconsistencies by element

| Element | Percentage of unique datasets (i.e., the total number of datasets with any disagreement, where each dataset is counted only once even if there is more than 1 type of disagreement, divided by the total number of datasets in a given round) | | | |
|---|---|---|---|---|
| | Training (5 datasets) | Round 1 (10 datasets) | Round 2 (5 datasets) | Round 3 (5 datasets) |
| Context, purpose, motivation | 0% | 0% | 0% | 0% |
| Requirements | 20% | 0% | 20% | 0% |
| Ethicality | 60% | 40% | 40% | 20% |
| Domain knowledge and data practices | 80% | 60% | 80% | 40% |
| Context awareness | 60% | 20% | 80% | 20% |
| Environmental footprint | 40% | 0% | 0% | 0% |
| Data collection | 40% | 0% | 20% | 0% |
| Data processing | 60% | 20% | 20% | 0% |
| Data annotation | 40% | 50% | 60% | 0% |
| Suitability | 0% | 0% | 0% | 0% |
| Representativeness | 40% | 60% | 40% | 0% |

| | | | |
|---|---|---|---|
| Authenticity | 100% | 80% | 40% | 60% |
| Reliability | 60% | 30% | 40% | 20% |
| Integrity | 80% | 30% | 40% | 60% |
| Structured documentation | 40% | 50% | 20% | 20% |
| Findability | 60% | 30% | 20% | 0% |
| Accessibility | 20% | 0% | 0% | 0% |
| Interoperability | 40% | 60% | 80% | 0% |
| Reusability | 20% | 40% | 20% | 20% |

### C.6        Review of structured documentation in NeurIPS datasets

In response to the range of evaluation outcomes for the 'structured documentation' dimension, we reviewed the current data practices reported by the dataset creators in more detail by analyzing the context documents provided with publications. Our review indicates that out of 25 datasets assessed, 6 lacked an accompanying context document. Of the 19 datasets with context documentation, we identified limitations that undermine their completeness and utility. This review highlights instances where modifications to standard datasheets or checklists by dataset creators lead to the omission of essential curation details. We document cases where the provided information was ambiguous or could not be independently verified, emphasizing the need for improved documentation standards to uphold the integrity of data curation processes.

Through this review, a recurrent challenge observed involves the difficulty that dataset creators face in clearly justifying their decision to omit a permanent identifier (such as a DOI) in their documentation. This omission often results in responses that appear evasive or misleading when addressing this aspect in the datasheet. However, it's important to recognize that there are legitimate constraints that may impede dataset creators from providing this information. An acknowledgment of such limitations is essential for transparency in data curation. For example, the fNIRS2MW dataset [62] transparently addresses the absence of a DOI. The authors openly state their ongoing exploration of viable options for obtaining a DOI, demonstrating a commitment to transparency in their data curation process. Such acknowledgments are critical in advancing data curation, as they provide insight into the practical challenges encountered and encourage a culture of openness and improvement.

Table 5: Structured documentation in NeurIPS datasets

| ID | Publication TItle | Reference | Application domain | Size | Documentation | Explanation of Strength | Explanation of Weakness |
|---|---|---|---|---|---|---|---|
| 1 | Programming Puzzles | [126] | Evaluation and benchmarking of program synthesis and AI coding capabilities | 397 instances of programming puzzles | Datasheet | The datasheet includes documentation on 7 data curation aspects, including motivation, composition, collection, preprocessing, usage, distribution, and maintenance. | We observed instances where the answers provided by the authors to the standard questions were ambiguous.<br><br>E.g. 1: When prompted, "Will the dataset be updated?", the authors respond in one word and mention "GitHub". This standard inquiry anticipates a definitive 'Yes' or 'No' along with an accompanying rationale. The authors' concise reply does not fulfill the expected detail, potentially leading to ambiguity about the procedures for dataset updates, such as the mechanism of corrections or the introduction of new data.<br><br>E.g. 2: When prompted, "How will the dataset be distributed? Does the dataset have a digital object identifier (DOI)?", the authors respond in one word and mention "GitHub". The latter question expects a definitive 'Yes' or 'No' along with an accompanying rationale. A more detailed response would help clarify the dataset's distribution method and the presence of a DOI, ensuring comprehensive understanding for users seeking to locate and cite the dataset. |
| 2 | Open Bandit Dataset and Pipeline: Towards Realistic and Reproducible Off-Policy Evaluation | [123] | Off-policy Evaluation (OPE), particularly in the context of batch bandit algorithms within e-commerce platforms. | ~26 M instances of user impressions | None Supplementary Material states that data documentation is available on the website (Page 22), but it is not findable. Assessed 16th Jan 2024. | - | - |
| 3 | SciGen: a Dataset for Reasoning-Aware Text Generation from Scientific Tables | [103] | Reasoning-aware data-to-text generation, specifically focusing | ~53 K instances of table-description pair | Datasheet | The datasheet includes documentation on 9 points, including motivation, composition, collection process, | In the datasheet, Page 4, under Distribution, the authors mention making DOI available after finalizing |

| | | | | | | | |
|---|---|---|---|---|---|---|---|
| | | | on generating descriptive text from tables in scientific articles. | | | preprocessing, usage, distribution, maintenance, metadata, and responsibility. | their GitHub. But, it is not available. Assessed 16th Jan 2024. |
| 4 | MOMA-LRG: Language-Refined Graphs for Multi-Object Multi-Actor Activity Parsing | [91] | Evaluation and benchmarking of video-language models (VLMs). | ~148 hours of annotated videos with several activity instances. | Checklist and Supplementary Material Checklist | | The checklist and Supplementary Material Checklist are misleading.<br><br>E.g. 1: The checklist mentions "Yes" to include limitations and negative social impacts of the work in the paper. But, none of these items are included.<br><br>E.g. 2: In the SM checklist, the authors mention, "Please see Section C" when they are highly encouraged to include a DOI. However, Section C does not provide relevant information to the readers. |
| 5 | CEDe: A collection of expert-curated datasets with atom-level entity annotations for Optical Chemical Structure Recognition | [60] | Optical Chemical Structure Recognition (OCSR), focusing on translating chemical images to molecular structures | ~700 K instances of chemical entity bounding boxes | Datasheet | The datasheet includes documentation on 6 points, including motivation, composition, collection process, uses, distribution, and maintenance while providing sufficient information across these points. | No documentation on the meta-data and responsibility sections of the datasheet. |
| 6 | LoveDA: A Remote Sensing Land-Cover Dataset for Domain Adaptive Semantic Segmentation | [136] | Remote sensing land-cover domain adaptive semantic segmentation. | ~ 6 K instances of images with several annotated objects. | None | - | - |
| 7 | RELLISUR: A Real Low-Light Image Super-Resolution Dataset | [1] | Evaluation and benchmarking of low-light enhancement and super-resolution image processing. | ~13 K instances of paired images | Datasheet (Non-standard) | The datasheet includes documentation on 9 points, including composition, collection process, uses, distribution, maintenance, and author statement. | The authors were selective in documenting the aspects posed by a standard data sheet, which poses a risk of missing information across several crucial aspects.<br><br>E.g. 1: Under "Data Use", a standard datasheet asks about the tasks for which the datasheet should be and shouldn't be used. The authors modified their datasheet by removing these individual questions and |

|  |  |  |  |  |  |  | documenting the information on the tasks for which the datasheet should be used while omitting the documentation on the tasks for which it should not be used.<br><br>These consistencies should have been addressed using a standard datasheet. |
|---|---|---|---|---|---|---|---|
| 8 | Measuring Mathematical Problem Solving With the MATH Dataset | [53] | Measuring and improving the mathematical problem-solving capabilities of machine learning models | ~12.5 K instances of mathematics problems and their solutions | Checklist (Non-standard) | The checklist includes information about licensing and information on the intended uses of the dataset. | This non-standard checklist does not document information on the composition, collection process, uses, and maintenance, among other essential items. |
| 9 | DGraph: A Large-Scale Financial Dataset for Graph Anomaly Detection | [61] | Graph Anomaly Detection (GAD), specifically in the finance domain. | ~3 M instances of nodes with 4 M edges | Checklist | The checklist documents information to ensure the reproducibility of benchmarking experiments and the use of existing or curating new assets. | The structure of the Y/N checklist does not motivate authors to include the crucial items from a data curation perspective. For instance, the authors of this work report "No" for the questions asking whether they included limitations of their work and any potential negative societal impacts without elaborating further.<br><br>The authors also write that the licensing information is present in Section 5. But, we observe that it is actually present in the appendix. |
| 10 | Change Event Dataset for Discovery from Spatio-temporal Remote Sensing Imagery | [93] | Detection of meaningful change events in satellite imagery, focusing on specific events like road construction and forest fires. | ~28 K instances of the change event in CaiRoad and ~2 K instances of the change event in CalFire. | Datasheet and checklist | The documentation on 7 points, including motivation, composition (incl. documentation on metadata), collection process, preprocessing, usage, distribution, and maintenance while providing sufficient information across these points. | No documentation on responsibility sections of the datasheet; all required questions answered |
| 11 | CAESAR: An Embodied Simulator for Generating Multimodal | [65] | Evaluation and benchmarking of models for recognizing | ~1 M and ~124 K instances of scenes obtained from the simulator in the | Datasheet | The datasheet provides documentation on the curation process covering 7 points, including data access, author | The documentation on data access mentions three datasets: CAESAR-XL, CAESAR-L, and CAESAR-S; however, the |

| | | | | | | | |
|---|---|---|---|---|---|---|---|
| | Referring Expression Datasets | | multimodal referring expressions. | CAESAR-XL and CAESAR-L, respectively. | | statement, motivation, composition, preprocessing, usage, and distribution. The questions on maintenance were answered but merged with those on distribution. | rest of the datasheet includes no information on CAESAR-S.<br><br>In the datasheet, there have been instances of not providing a direct answer to a standard question. For instance, on Page 5, when asked, "How will the dataset be distributed (e.g., tarball on website, API, GitHub)? Does the dataset have a digital object identifier (DOI)?" The authors mention, "We will share the dataset download links (splitted zip files) after receiving a data use agreement (DUA)." In their response, the authors ignore the latter question by not providing a direct answer. |
| 12 | GLOBEM Dataset: Multi-Year Datasets for Longitudinal Human Behavior Modeling Generalization | [139] | Evaluation and benchmarking of models for longitudinal behavior modeling and depression detection. | Several instances of 497 unique participants across four years. | Checklist and data document (Nonstandard) | The checklist documents information to ensure the reproducibility of benchmarking experiments and the ethical conduct involved in recruiting human participants.<br><br>The data document covers information on author contributions, hosting (including long-term preservation), licensing, maintenance, meta-data, usage, distribution, privacy, and ethics. | The checklist is misleading. The authors state that they discuss potential negative societal impacts of their work in the discussion section (Section 6). Yet, there is no such discussion. The authors mention the limitation of their work from a machine-learning lens; however, a discussion on potential negative aspects is missing.<br><br>Although the authors curated a new dataset, they mentioned 'Not Applicable' under all the necessary items under the new dataset curation subsection (Bullet point 4).<br><br>The use of a nonstandard data document leads the authors to selectively document information on dataset curation. On the one hand, the data document goes beyond documenting long-term preservation; on the other hand, it does not explicitly include information on data composition. A standard datasheet asks the authors to document the number of instances and their composition. |

| 13 | ClimateSet: A Large-Scale Climate Model Dataset for Machine Learning | [70] | To support machine learning-based climate model emulation, downscaling, and prediction tasks. | Several instances obtained by combining 36 climate models from the Input4MIPs and CMIP6 archives. | None | - | - |
|----|----|----|----|----|----|----|----|
| 14 | BubbleML: A Multiphase Multiphysics Dataset and Benchmarks for Machine Learning | [50] | Evaluation and benchmarking of models to detect phase change phenomena. | 6 instances of boiling simulations. | Datasheet | The datasheet documents information on motivation, composition, collection, preprocessing, usage, distribution, maintenance, reproducibility, and data format. | None identified. |
| 15 | DataComp: In search of the next generation of multimodal datasets | [38] | Evaluation and benchmarking for multimodal machine learning tasks like image recognition and natural language processing. | ~12.8 B instances of image url-text pairs | Datasheet | The documentation covers 7 aspects of data curation, including motivation, composition, collection, preprocessing, usage, distribution, and maintenance. | None identified. |
| 16 | The CPD Data Set: Personnel, Use of Force, and Complaints in the Chicago Police Department | [59] | Facilitation of machine learning research in policing, specifically police behavior, misconduct prediction, and violence. | Several instances covering 35 K officers, 730 K award request records, 194 K salary records, 108 K unit assignments, 109 K complaints, and 11 K tactical response reports. | Checklist and datasheet | The checklist documents aspects to ensure reproducibility and the use of existing or curating new assets.<br><br>The datasheet covers information on motivation, composition, collection, preprocessing, usage, distribution, and maintenance. | No documentation on the responsibility section of the datasheet. |
| 17 | The Tufts fNIRS Mental Workload Dataset & Benchmark for Brain-Computer Interfaces that Generalize | [62] | Developing models that accurately classify mental workload intensity levels. | 30-60 minute instances of functional near-infrared spectroscopy (fNIRS) from 68 human participants. | Datasheet | The datasheet covers documentation on motivation, composition, collection, preprocessing, usage, distribution, and maintenance. | No documentation on the responsibility section of the datasheet. |
| 18 | How Would The Viewer Feel? Estimating Wellbeing From Video Scenarios | [95] | Detecting the emotional response and subjective well-being elicited by video content. | 60 K instances of manually annotated videos for emotional response and subjective well-being. | X-Risk Sheet | The X-Risk Sheet documents potential existential risks from future AI systems to facilitate better understanding and effective mitigation strategies. | The documentation does not cover information on fundamental data curation practices, such as the composition, collection process, usage, and maintenance, among other essential items. |

| 19 | The RefinedWeb Dataset for Falcon LLM: Outperforming Curated Corpora with Web Data Only | [115] | Training and benchmarking large language models | 10 B instances of text-only documents, corresponding to single web pages. | Checklist, Datasheet, and Model Card (Falcon-RW) | By including a Model Card, the authors document their model's architecture and training, ethical considerations, potential biases, and performance metrics.

The datasheet further documents 7 aspects of data curation, including motivation, composition, collection, preprocessing, usage, distribution, and maintenance.

Lastly, the checklist documents information to ensure the reproducibility of benchmarking experiments and the use of existing or curating new assets. | None identified. |
| 20 | Stanford-ORB: A Real-World 3D Object Inverse Rendering Benchmark | [77] | Evaluation and benchmarking for inverse rendering methods for 3D objects. | Several instances of 14 real-world 3D objects | None | - | - |
| 21 | FS-Mol: A Few-Shot Learning Dataset of Molecules | [130] | Modeling quantitative structure-activity relationships (QSAR) in early drug discovery to identify novel active molecules. | 233 K unique compounds across a total of 5120 separate assays. | None | - | - |
| 22 | Evaluating Out-of-Distribution Performance on Document Image Classifiers | [79] | Evaluating the performance of document classifiers on out-of-distribution data. | 4 K out-of-distribution document images: 1 K (RVL-CDIP-N) and 3 K (RVL-CDIP-O) | None | - | - |
| 23 | Dungeons and Data: A Large-Scale NetHack Dataset | [48] | Reinforcement learning in the context of gaming, specifically using game-playing | 10 B state transitions from 1.5 million human games and 3 B state transitions from 100 K | Checklist | The checklist records information to ensure experiment reproducibility and the use of existing or newly curated assets. While the | None identified. |

| | | | trajectories to improve decision-making algorithms. | games played by a symbolic bot. | | checklist allows for Yes/No responses, authors provide appropriate justifications where necessary. | |
|---|---|---|---|---|---|---|---|
| 24 | VisAlign: Dataset for Measuring the Alignment between AI and Humans in Visual Perception | [82] | Measuring AI-human visual alignment in image classification tasks. | 14 K instances of an image and its corresponding label | Datasheet | The datasheet covers documentation on motivation, composition, collection, preprocessing, usage, distribution, and maintenance. | The datasheet omits answers to several questions, including the ones related to potential harm or ethics. It also lacks detailed information on composition, creation, and preprocessing, which can be found in the main paper. |
| 25 | American Stories: A Large-Scale Structured Text Dataset of Historical U.S. Newspapers | [26] | Improving historical document digitization and language model training through structured newspaper data analysis. | 1 B content bounding boxes from historical U.S. newspaper scans. | Datasheet | The datasheet covers documentation on motivation, composition, collection, preprocessing, usage, distribution, and maintenance. | None identified. |



\end{document}